\documentclass[journal=aesccq,manuscript=article]{achemso}

\usepackage[version=3]{mhchem} 
\usepackage{subcaption}
\usepackage[title]{appendix}

\author{Francesco Benedetti}
\affiliation[Sapienza]
{Department of Chemistry, Sapienza University of Rome, P.le Aldo Moro 5, 00185 Rome, Italy}
\author{Mauro Satta}
\affiliation{Institute of the Study of Nanostructured Materials (ISMN-CNR), Department of Chemistry, Sapienza University of Rome, P. le Aldo Moro 5, 00185 Rome, Italy}
\author{Tommaso Grassi}
\affiliation[mpe]
{Center for Astrochemical Studies, Max-Planck-Institut f\"ur Extraterrestrische Physik, Giessenbachstr. 1, D-85748 Garching, Germany}
\author{Stefan Vogt-Geisse}
\affiliation[Conce]
{Departamento de Físico-Química, Facultad de Ciencias Químicas, Universidad de Concepción, 4070386 Concepción, Chile}
\author{Stefano Bovino}
\affiliation[Sapienza]
{Department of Chemistry, Sapienza University of Rome, P.le Aldo Moro 5, 00185 Rome, Italy}
\alsoaffiliation[Second University]
{Departamento de Astronomía, Facultad Ciencias Físicas y Matemáticas, Universidad de Concepción Av. Esteban Iturra s/n Barrio
Universitario, Casilla 160, Concepción, Chile}
\email{stefano.bovino@uniroma1.it}
\alsoaffiliation[Second University]
{INAF, Osservatorio Astrofisico di Arcetri, Largo E. Fermi 5, 50125 Firenze, Italy}
\alsoaffiliation[CNR]
{Institute of Structure of Matter (ISM-CNR), Consiglio Nazionale delle Ricerche, 34149 Basovizza, Italy}

\title{CO diffusion on interstellar amorphous solid water: a computational study}


\begin{document}

\begin{tocentry}

    \includegraphics[scale=0.195]{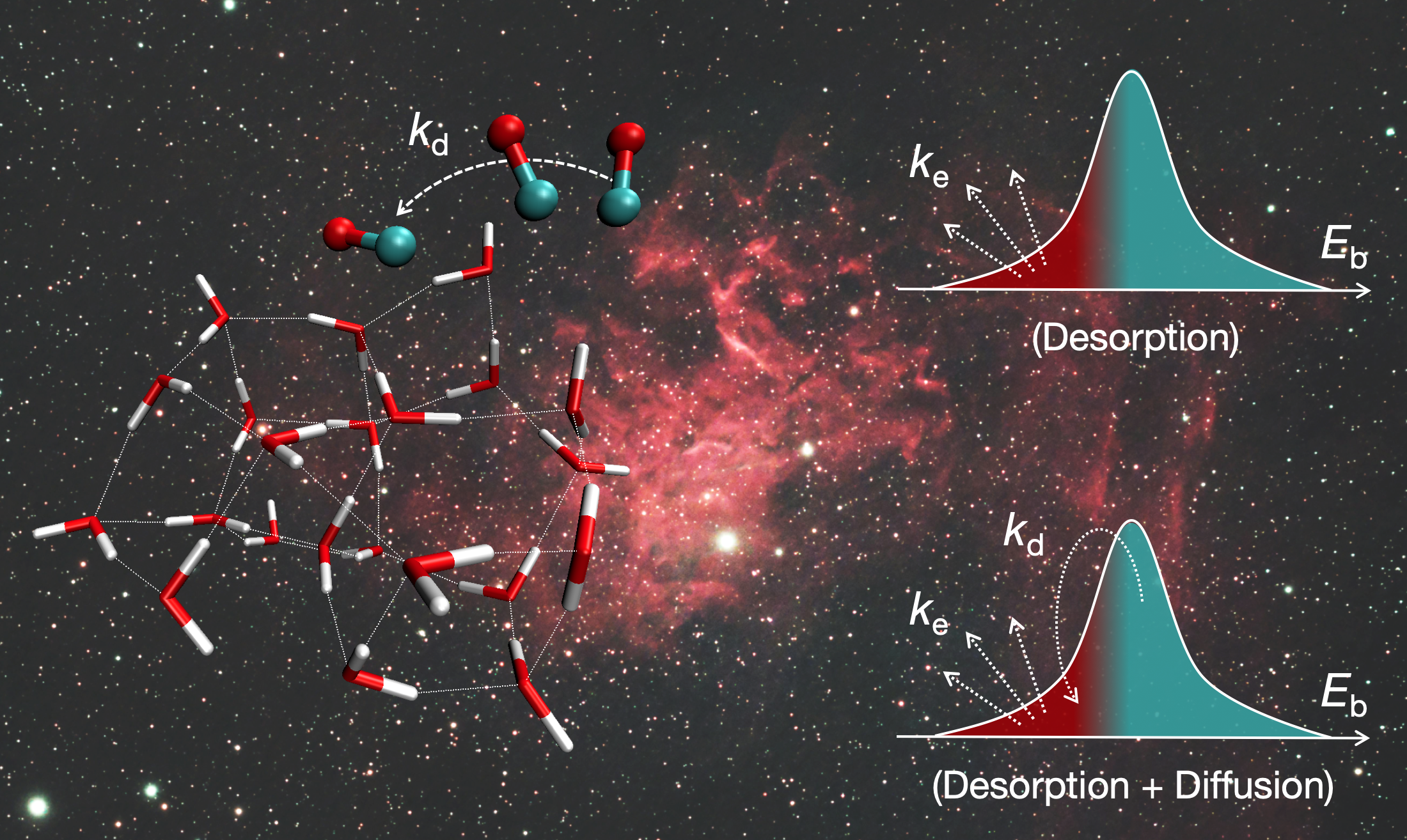}

\end{tocentry}

\newpage

\begin{abstract}
Surface chemistry on interstellar dust grains is recognized as a central component in astrochemical models, representing a plausible formation route for many of the observed complex molecular species. However, key parameters governing interstellar surface chemistry, such as diffusion energy barriers, remain poorly constrained. In particular, surface diffusion constitutes a fundamental step for the synthesis of complex organic molecules and plays a crucial role in understanding the desorption process.
In this paper, the diffusion dynamics of carbon monoxide (CO) on amorphous solid water (ASW) surfaces, representative of interstellar ices, is modeled with quantum-chemical methods. Employing a representative ensemble of water clusters, each made by 22 molecules, diffusion energy barriers between the binding sites are computed using Density Functional Theory. Diffusion rate coefficients are then determined by applying the harmonic approximation of Transition State Theory.
The results, in agreement with experimental studies, revealed a wide distribution of diffusion energies. This reflects the intrinsic topological heterogeneity of ASW surfaces, and highlights how surface mobility significantly influences CO's desorption dynamics and, as a consequence, surface-mediated reactivity in interstellar environments. We show that key parameters commonly employed in astrochemical models, like the ratio between binding and diffusion energy, should be carefully revised.
\end{abstract}

\newpage 

\section{Introduction}

Star- and planet-formation processes occur in the dense cores of interstellar molecular clouds~\cite{van1998chemical, jorgensen2020astrochemistry}, characterized by specific physical conditions ($T < 20$~K, number densities~$\sim 10^4$~cm$^{-3}$)~\cite{van2017astrochemistry}, that allow dust grains to be covered by a thick ice mantle mostly of amorphous solid water (ASW)~\cite{fraser2002laboratory, draine2003interstellar} as well as other molecules such as CO, NH$_3$, CO$_2$, and CH$_4$~\cite{boogert2015observations}. More recently, observations with the James Webb Space Telescope (JWST) have revealed these ices with unprecedented details, providing further insights into their composition and evolution in dense clouds~\cite{mcclure2023ice}. 
\\
In these extreme conditions, where most of the gas-phase reaction channels are suppressed, chemistry takes place on the interstellar ice layers~\cite{cuppen2017, cuppen2024}. Molecules from the gas phase accrete on the icy surfaces, where reactive encounters promote the formation of new molecules. Since the diffusive Langmuir-Hinshelwood mechanism is the primary process involved in interstellar surface chemistry~\cite{cuppen2017}, the mobility of adsorbed species plays a crucial role in the synthesis of the complex organic molecules (COMs), since it determines the encounter probability of the reactants, both through thermally-activated hops and quantum tunneling~\cite{hama2013surface}. 
Diffusion is also important for the processes happening in the bulk of the ice layers. When the ice mantle exceeds the thickness of a single monolayer, the molecules below the surface need to diffuse through the bulk of the ice, to reach the most external layer and desorb towards the gas phase. This interplay between desorption and bulk diffusion has been investigated in astrophysical scenarios~\cite{ghesquiere2015diffusion}.
\\
Despite their relevance, diffusion processes on icy surfaces have been poorly investigated. This is due to the difficulties associated with experimental measurements of diffusion rate coefficients~\cite{mate2020diffusion, minissale2016direct}, often based on the indirect determination of other parameters, thus yielding uncertain results. As a consequence, surface diffusion remains one of the least characterized processes in astrochemical models. To overcome this limitation, in the majority of gas-grain chemical models, the formalism introduced by Hasegawa et al.~\cite{hasegawa1992models} is adopted, where the reaction rate is corrected by a factor that includes both thermal hopping and quantum tunneling effects. The thermal hopping rate is described by the Arrhenius-like expression
\begin{equation}
\label{diff_arrhenius}
    k_{\text{d}} = \nu_{\text{d}} \exp{\bigg(-\frac{E_{\text{d}}}{k_{\text{B}}T}\bigg)} \,,
\end{equation}
where $\nu_\mathrm d$ corresponds to a characteristic vibrational frequency (typically $10^{12}-10^{13}$~s$^{-1}$) and $E_\mathrm d$ to the diffusion energy barrier. In absence of direct measurements, $E_\mathrm d$ is usually approximated as a fixed fraction, $0 < \alpha \leq 1$, of the adsorption energy $E_\mathrm b$, with the relation $E_{\text{d}} = \alpha E_{\text{b}} \,,$ a very strong and purely empirical assumption.
\\
In reality, it is reasonable to consider a variable $\alpha$, depending on the molecular species, the surface structure and composition, and its coverage. However, this parameter is commonly treated as a constant in astrochemical models, employing values between 0.3 and 0.7~\cite{garrod2006formation, cuppen2009microscopic, ruffle2000new,semenov2010chemistry,jin2020formation,walsh2014complex}, with most recent modeling efforts tending to adopt values closer to the lower end.
In addition to the energy barrier, another key parameter in the computation of diffusion rate coefficients is the pre-exponential factor $\nu_\mathrm d$ (see Eq.~\ref{diff_arrhenius}). In particular, it is usually assumed to be equal to the desorption’s pre-exponential factor, increasing the overall uncertainty~\cite{chen2024astrochemical}. As a consequence, the approximation commonly employed in astrochemical models does not reflect the reality of interstellar surfaces, particularly in the case of amorphous ices, where the physico-chemical properties can change significantly between the various binding sites~\cite{bovolenta2024depth}. For a deeper discussion and a comparison between theoretical and experimental determination of the pre-exponential factor, see the review of Minissale et al.\cite{minissale2022_review}.
\\

Experimental investigations aimed at characterizing molecular diffusion on interstellar ice analogues, although limited in number, have provided several pieces of information. Most of these works focused on atomic~\cite{minissale2016direct,tsuge2023surface} diffusion, and more specifically on hydrogen~\cite{Watanabe2010,kimura2018,hama2012mechanism,matar2008mobility,kuwahata2015signatures}, on different substrates. One of the first attempts at constraining molecular diffusion was conducted by \"Oberg et al.~\cite{oberg2009quantification}, in segregation experiments of H$_2$O:CO mixtures. Analyzing segregation kinetics employing Reflection-Absorption Infrared Spectroscopy (RAIRS) techniques, they derived a diffusion energy barrier for CO equal to $0.60 \pm 0.20$~kcal\,mol$^{-1}$ ($300 \pm 100$~K), consistent with an Arrhenius-like behaviour. 
\\
A more direct approach, involving Fourier Transform Infrared spectroscopy, was employed by Mispelaer et al.~\cite{mispelaer2013diffusion}, investigating the surface mobility of several molecules (CO, HNCO, H$_2$CO, and NH$_3$) on porous ASW ice over the temperature range between 35 and 140~K. In particular, the reported diffusion energy for CO was of $0.24 \pm 0.36$~kcal\,mol$^{-1}$ ($120 \pm 180$~K).
\\
Among the most relevant experimental investigations is the work of Lauck et al.~\cite{lauck2015co} where, through a combination of infrared analyses and Fick's law–based theoretical modeling, they determined a diffusion energy for CO on ASW ices of $0.31 \pm 0.02$~kcal\,mol$^{-1}$ ($158 \pm 12$~K).
\\
A significantly higher value of $0.97 \pm 0.02$~kcal\,mol$^{-1}$ ($490 \pm 12$~K) was reported by He et al.~\cite{he2018measurements}, derived from a RAIRS analysis.
\\
More recently, an ultra-high-vacuum transmission electron microscope was used to directly monitor the deposition and mobility of CO and CO$_2$ molecules on ASW substrates~\cite{kouchi2020direct}, deriving a diffusion energy barrier for CO equal to $0.70 \pm 0.10$~kcal\,mol$^{-1}$ ($350 \pm 50$~K). This method was later extended to a series of adsorbates~\cite{furuya2022diffusion}, revealing that the relation between diffusion and desorption barriers is not universal, as the $E_\mathrm d / E_\mathrm b$ ratio strongly depends on the specific molecule.
\\
However, experimental studies face some major limitations. Firstly, diffusion and desorption are often coupled, making it difficult to isolate the contribution of each process. In addition, ice structural changes happening during the measurements, such as pore collapse~\cite{cazaux2015pore} or surface reorganization~\cite{he2019effective}, can affect molecular mobility, introducing further uncertainties.
\\
It is also worth noting that such experiments do not consistently replicate interstellar conditions. In an Earth-based laboratory, adsorbate-adsorbate interactions tend to dominate, leading to a concentration-driven diffusion. On the other hand, in the typical interstellar conditions, where the surface coverage is extremely low, adsorbate–adsorbent interactions prevail and adsorbate–adsorbate interactions become negligible~\cite{he2018measurements}. Specifically, most experimental techniques do not measure microscopic tracer diffusion, but rather macroscopic transport driven by concentration (chemical potential) gradients. As a result, deriving the parameters appearing in Eq.~\ref{diff_arrhenius} requires modeling assumptions linking the measured macroscopic flux to atomic-scale hops, which introduce significant and not easily quantifiable uncertainties.
This fundamental difference limits the direct applicability of laboratory-derived diffusion parameters to astrochemical models. 
\\

Theoretically, only a limited number of studies addressed the problem~\cite{asgeirsson2017long,senevirathne2017hydrogen,zaverkin2022neural}, mainly via kinetic Monte Carlo (KMC) methods~\cite{andersen2019practical,cuppen2013kinetic}. In a study by Karssemeijer et al.~\cite{karssemeijer2012long}, a combination of on- and off-lattice KMC simulations was performed to model CO diffusion on crystalline ice, deriving an energy barrier of $1.15 \pm 0.02$~kcal\,mol$^{-1}$ ($580 \pm 12$~K).
In a following study, Karssemeijer et al.~\cite{karssemeijer2013dynamics} combined KMC simulations with isothermal desorption experiments to investigate CO diffusion dynamics on ASW surfaces. With the pores being partially filled, they report a convergence to a energy barrier of $1.52 \pm 0.46$~kcal\,mol$^{-1}$ ($766 \pm 232$~K), a value higher than the experimental estimates of about $0.60 \pm 0.35$~kcal\,mol$^{-1}$ ($302 \pm 174$~K), likely due to the presence of CO–CO interactions. This study also confirms how diffusion on crystalline ice is significantly faster, with a barrier being roughly 50\% lower than the one on amorphous substrates.
\\
From the collection of studies discussed above, a great variability emerges in the values of the energy barriers, reflecting the results' sensitivity to both experimental conditions and modeling assumptions. This dispersion highlights the complexity of diffusion processes in interstellar ices and the difficulties in obtaining consistent and reliable data. Improving our understanding of diffusion mechanisms and providing reliable theoretical values for fundamental parameters is therefore a key priority in the development of realistic models for reactivity in interstellar environments. In this work, we address this gap by performing quantum-chemical calculations of the diffusion energy barriers for CO molecules adsorbed on interstellar ice analogues.
\\

Although the influence of CO's diffusion energies remains relatively unexplored in astrochemical models, recent studies confirm that their magnitude significantly affects the formation kinetics of various molecules~\cite{acharyya2022understanding}. Therefore, understanding CO's diffusion dynamics on ASW surfaces remains a key priority in solid-state astrochemistry, as it directly affects the predictive capabilities of chemical models in different interstellar environments. 
A more comprehensive and detailed discussion of astrophysically-relevant diffusion processes can be found in the recent contribution of Ligterink et al.~\cite{ligterink2025molecular} and references therein.
\\
In the following Sections, we describe in detail the theoretical calculations employed in the present investigation, based on Density Functional Theory (DFT), and we suggest possible implications of the newly derived parameters on surface kinetics in astrophysically relevant scenarios.

\section{Methodology and computational details}\label{sec_methods}
To investigate CO's surface mobility, we adopt the cluster approach, taking as reference the ASW systems composed of 22 water molecules developed by Bovolenta et al.~\cite{bovolenta2020high,bovolenta2022}. These clusters have been generated through classical molecular dynamics (MD) simulations, employing the TIP3P model for water, and subsequently amorphized at 300~K by means of ab initio molecular dynamics (AIMD) simulations, carried out at BLYP/def2-SVP level of theory. 100 uncorrelated configurations were extracted from the AIMD trajectory and then cooled to 10~K by annealing, in order to reproduce conditions close to the interstellar ones. The resulting structures were finally grouped according to geometric similarity, with a Root Mean Square Deviation (RMSD) threshold of 0.4~\AA, selecting the 20 most representative configurations. The binding sites identified by Bovolenta et al.~\cite{bovolenta2022} represent the basis of the present investigation, since the diffusion network was determined by calculating the minimum energy paths (MEPs) between the different energy minima, after a selection of the most representative pairs of adsorption sites (as explained in the following Sections).
Additionally, in previous studies it has been shown how increasing the cluster size has only a minor effect on the resulting binding energy distributions~\cite{bovolenta2020high}, indicating that the essential local structural features of ASW are already well captured at this scale.
For these reasons, we expect cluster-size effects on the computed diffusion barriers to be limited.
In this work, we adopt a DFT-based computational protocol for the structural and energetic analysis of ASW clusters. The theoretical details regarding each step of the procedure are explained in the following Sections.
\subsection{Geometrical optimizations and vibrational analysis}
The M06-2X functional~\cite{zhao2008m06}, known for its high accuracy in the treatment of weakly bound systems~\cite{hohenstein2008assessment}, was used for the geometrical optimizations and vibrational analysis. Indeed, M06-2X has been known to outperform some of the most popular functionals and methods in the treatment of hydrogen-bonded systems. For instance, it has been shown to offer improved accuracy over B3LYP in the modeling of protonated methanol clusters~\cite{fifen2013solvation} and to have a better performance with respect to the MP2 method in the description of hydrogen-bonding networks in ammonia clusters~\cite{malloum2015structures}. In addition, previous studies have highlighted its applicability to small water clusters, structurally analogous to the systems involved in this work. In a systematic analysis conducted by Howard et al.~\cite{howard2015assessing}, on water clusters (H$_2$O)$_n$ with $n = 2$–6, M06-2X has shown a good performance in the prediction of harmonic vibrational frequencies, exhibiting a low shift with respect to reference CCSD(T)/CBS data. Although small water clusters have been extensively benchmarked~\cite{chedid2021energies, dahlke2005improved, dahlke2008assessment}, medium-sized clusters ($10 \leq n < 20$) introduce greater difficulties, due to the complexity of the potential energy surface (PES). In this context, Malloum et al.~\cite{malloum2019structures} showed that, through an extensive conformational sampling followed by geometrical optimizations at M06-2X/6-31++G(d,p) level of theory, it is possible to identify candidates for global minima for neutral clusters up to $(\text{H}_2\text{O})_{30}$, in good agreement with reference CCSD(T)/CBS data. This evidence supports the reliability of the M06-2X functional and justifies its use in the present investigation.
\\
Although M06-2X was chosen for its robustness in the structural and vibrational analysis, a different functional was employed to refine the energetic description of the diffusive events. In this context, the MPWB1K functional~\cite{zhao2004hybrid} was selected, given its accuracy in the treatment of systems held together by non-covalent interactions~\cite{zhao2005benchmark, dkhissi2007performance}. The complementary use of these functionals ensures a robust DFT framework for the characterization of the surface diffusion events.
\\

All DFT calculations were performed using the \textsc{orca}~6.0 software package~\cite{neese2020orca}. Structural optimizations and vibrational analysis were performed at M06-2X/def2-TZVP level of theory, including Grimme’s D3 dispersion correction with the zero-damping scheme~\cite{grimme2010consistent}, in the gas phase and without any symmetry constraints. Two different optimization protocols were adopted, depending on the nature of the stationary point. For local minima, the Broyden–Fletcher–Goldfarb–Shanno (BFGS) algorithm~\cite{schlegel2003exploring} was employed in redundant internal coordinates, with an initial Hessian estimated using the Almloef model and a maximum of 204 optimization steps. For first-order saddle points, the Bofill update algorithm~\cite{bofill1994updated} was used, with a precomputed Hessian as an initial guess and a maximum of 500 optimization steps. The main convergence parameters adopted for both optimization protocols are summarized in Tab.~\ref{tab:convergence}.

\begin{table}[htbp]
    \centering
    \caption{\label{tab:convergence}Convergence thresholds employed for the geometrical optimizations of local minima and first-order saddle points.}
    \begin{tabular}{l c c}
        \hline
         \textbf{Parameter} & \textbf{Local minima}  & \textbf{Saddle points} \\
        \hline
        Energy change & $1.0 \times 10^{-6}$~E$_\mathrm{h}$ & $5.0 \times 10^{-6}$~E$_\mathrm{h}$ \\
        Maximum gradient & $1.0 \times 10^{-4}$~E$_\mathrm{h}$\,bohr$^{-1}$ & $3.0 \times 10^{-4}$~E$_\mathrm{h}$\,bohr$^{-1}$ \\
        RMS gradient & $3.0 \times 10^{-5}$~E$_\mathrm{h}$\,bohr$^{-1}$ & $1.0 \times 10^{-4}$~E$_\mathrm{h}$\,bohr$^{-1}$ \\
        Maximum displacement & $1.0 \times 10^{-3}$~bohr & $4.0 \times 10^{-3}$~bohr \\
        RMS displacement & $6.0 \times 10^{-4}$~bohr & $2.0 \times 10^{-3}$~bohr \\
        \hline
    \end{tabular}
\end{table}

The nature of the stationary points was later confirmed by harmonic vibrational frequency calculations: local minima were identified by the absence of imaginary frequencies, while transition states exhibited a single imaginary frequency along the diffusion coordinate. Vibrational frequencies were also used to compute the zero-point energy (ZPE) corrections to the binding energies and the molecular partition functions, which are required for the subsequent rate calculation.
\subsection{Diffusion pathway analysis}
The diffusion pathway analysis was conducted using the Nudged Elastic Band (NEB) method~\cite{asgeirsson2021nudged}, as implemented in the \textsc{orca}~6.0 software. Calculations for each image were performed at M06-2X-D3/def2-TZVP level of theory and, additionally, the NEB-TS variant was employed to refine the geometry of the saddle point once the MEP had been approximately located, ensuring the accurate identification of transition states along the diffusion coordinate. Each NEB calculation involved 10 intermediate images, for a total of 12 images including the initial and final minima. The MEP estimate was carried out using the improved tangent scheme~\cite{henkelman2000improved} and the CI-NEB variant~\cite{henkelman2000climbing} was activated once the force on the images dropped below the threshold of $2.0 \times 10^{-2}$~E$_\mathrm{h}$\,bohr$^{-1}$. Energy-weighted elastic forces were applied between images, with spring constants ranging from 0.0100 to 0.1000 E$_\mathrm{h}$\,bohr$^{-2}$ and distributed according to the distance between adjacent images. Pathway convergence was assessed for each image employing the L-BFGS optimization algorithm~\cite{sheppard2008optimization}, with a maximum of 500 iterations. Convergence thresholds were set to $2.0 \times 10^{-2}$~E$_\mathrm{h}$\,bohr$^{-1}$ for the maximum force and to $1.0 \times 10^{-2}$~E$_\mathrm{h}$\,bohr$^{-1}$ for the RMS force on the entire band. Stricter criteria were applied to the climbing image: $2.0 \times 10^{-3}$~E$_\mathrm{h}$\,bohr$^{-1}$ for the maximum force and $1.0 \times 10^{-3}$~E$_\mathrm{h}$\,bohr$^{-1}$ for the RMS force.
\\
To refine the energetic description of each diffusion event, single-point energy calculations were performed on the previously optimized structures (for the energy minima and for the transition states) at MPWB1K-D3BJ/def2-TZVP level of theory. The resulting energies, combined with ZPE corrections obtained from the vibrational analysis, were used to compute the ZPE-corrected diffusion energy barriers $\Delta E^{\ddagger}$ as
\begin{equation}
\label{diff_barriers}
    \Delta E^{\ddagger} = (E^{\ddagger} + \text{ZPE}^{\ddagger}) - (E^{i} + \text{ZPE}^{i}) \,,
\end{equation}
where the superscripts ${\ddagger}$ and ${i}$ represent  the transition state and the initial minimum, respectively. Employing this protocol, ZPE-corrected diffusion barriers were calculated for all pairs of sites selected for the NEB analysis.

\subsection{Diffusion and evaporation rate coefficients}
Once the energy minima and saddle points have been properly characterized and the diffusion energy barriers estimated, it is possible to compute the rate coefficients associated with each diffusion event. Diffusion rates are calculated using the harmonic approximation of Transition State Theory (hTST)~\cite{truhlar1996current}, incorporating the entire roto-vibrational contribution to the molecular partition functions. For every pair of binding sites, the diffusion rate coefficient $k_{\rm d}$ is computed as a function of temperature using the relation
\begin{equation}
\label{k_diff}
    k_{\text{d}}(T) = \frac{k_{\text{B}} T}{h}\sqrt{\frac{A^{i}B^{i}C^{i}}{A^{\ddagger}B^{\ddagger}C^{\ddagger}}} \frac{q^{\ddagger}_{\text{vib}}(T)}{q^{i}_{\text{vib}}(T)} e^{-\frac{\Delta E^\ddagger}{k_{\text{B}}T}} \,,
\end{equation}
where $A$, $B$ and $C$ represent the rotational constants of the initial ($i$) and transition states ($\ddagger$), and $q_{\rm{vib}}$ denotes the vibrational partition functions evaluated at the respective geometries. The exponential term includes the ZPE-corrected diffusion barrier, $\Delta E^\ddagger$. This protocol is applied systematically to every pair of binding sites, computing the rate coefficients for both the forward and reverse processes, thus providing a complete kinetic picture of bidirectional mobility. Additionally, the rate coefficients are determined over a temperature range between 10 and 100 K, consistent with the typical conditions of interstellar ices. 
\\

Furthermore, the desorption process was also taken into consideration. For each energy minima, the desorption rate coefficient $k_{\rm e}$ is estimated employing an analogous hTST-based approach
\begin{equation}
\label{des_rate}
    k_{\text{e}}(T) = \frac{k_{\text{B}} T}{h}\frac{Q^\text{CO}(T)Q^\text{cluster}(T)}{Q^{\text{CO+cluster}}(T)}e^{-\frac{E_\text{b}}{k_{\text{B}}T}} \,,
\end{equation}
where $Q^\text{CO}$ and $Q^\text{cluster}$ denote the ro-vibrational partition functions of the isolated CO molecule and of the ASW cluster, respectively, while $Q^{\text{CO+cluster}}$ represents the ro-vibrational partition function of the bound system (see Appendix~\ref{app_A} for more details on the computation of partition functions). The desorption energy $E_{\rm b}$ is equal to the ZPE-corrected binding energy of the corresponding minimum.
\\

The relevance of quantum tunneling effects in low-temperature diffusion decreases significantly for heavy species, although there is some evidence that atoms as heavy as oxygen may overcome the energy barrier via tunneling under specific conditions~\cite{congiu2014efficient}. Therefore, for the present investigation, the tunneling contribution is neglected due to the relatively high mass of the CO molecules.
\section{Results and discussion}\label{sec_results}
In this Section, we present the main results obtained for the diffusion of CO molecules on the ASW clusters, focusing on the computed energy barriers and rate coefficients. The analysis of the following data provides valuable insights into the role of molecular mobility which are relevant for interstellar surface reactivity. Before discussing the DFT results in detail, we describe the topological analysis used to classify the binding sites available on the ASW cluster surfaces. 

\subsection{Selection of binding sites}
The first stage concerned the mapping of the spatial distribution of the binding sites on the ASW clusters, identified as the center-of-mass positions of the CO molecules in their equilibrium configurations. Once the binding sites are identified, the  distances between every pair of sites are calculated, as they represent the basis for defining potential diffusion paths. The initial dataset consists of 1895 pairs of sites, with a mean distance of 9.52~\AA, and it was progressively refined by applying a series of geometrical and topological criteria, aimed at removing redundant or physically implausible connections. Indeed, not all neighboring sites are accessible to a CO molecule through a single diffusive step, particularly in the presence of intermediate obstacles or alternative pathways. The different stages of the protocol employed for the selection of site pairs are the following:
\begin{enumerate}
    \item \textbf{Energetic filter:} site pairs characterized by nearly identical values of $E_\mathrm b$ are considered redundant and removed, reducing the risk of reproducing equivalent pathways. Given two pairs, for example sites ($i,j$) and ($k,l$), the differences in binding energy between the respective endpoints $(\Delta E_{\text{b}}^{i-k}$ and $\Delta E_{\text{b}}^{j-l}$), as well as the cross-combinations ($\Delta E_{\text{b}}^{i-l}$ and $\Delta E_{\text{b}}^{j-k}$), are calculated. If both differences are below a threshold of 0.02 kcal\,mol$^{-1}$, one of the two pairs is discarded.
    \item \textbf{Angular filter:} it concerns cases in which a third binding site is almost collinear to a given site pair. In such scenarios, the diffusion process will happen more likely through the intermediate site rather than directly between the two original sites, and a direct migration would not occur. For each cluster, a Delaunay triangulation~\cite{virtanen2020scipy} is applied to the set of binding sites and, for every pair, the angle $\theta$ formed with a  third site is computed. If this angle is greater than a predefined threshold $\theta_\mathrm{thr}$, the original site pair is considered geometrically valid for direct diffusion. Otherwise, the pathway is discarded under the assumption that diffusion preferentially occurs through the intermediate site. In this study, a threshold of $\theta_{\text{thr}} = 15^\circ$ was employed, which reduces the number of pairs to 995, with an average distance of 8.70~\AA. A smaller threshold would artificially exclude a large number of connections, even in cases without true collinearity, while a larger value would retain many quasi-collinear configurations, which represent an issue for the subsequent diffusion pathway estimation. 
    \item \textbf{Topological filter:} it aims to exclude connections that require the CO molecule to cross the interior of the ASW cluster, as this does not represent a proper surface diffusion event. To identify these cases, a triangulated mesh enclosing the outermost water molecules of each cluster is constructed, effectively representing the accessible surface area. The segments connecting each pair of binding sites are then tested for possible intersections with the mesh. If the segment intersects it, the corresponding pathway is considered as passing through the bulk of the ice, and the pair is discarded. Otherwise, pairs whose trajectory does not intersect the surface are considered valid for surface diffusion. This procedure further restricts the set to 423 valid pairs, with an average distance of 6.68~\AA.
    \item \textbf{Cutoff and final selection:} A final, distance-based criterion is introduced, excluding all site pairs separated by more than 5~\AA~in distances. This choice is consistent with the assumptions typically adopted in astrochemical models, where inter-site distances are often taken to be on the order of $\sim 3$~\AA~\cite{hocuk2015interplay,hama2012mechanism,mate2020diffusion}. By applying this threshold, 124 valid pairs are obtained, with an average distance of 3.64~\AA, in good agreement with the expected values. This choice, in addition to being consistent with the typical length scale of diffusion steps at low temperatures, is also motivated by the nature of the initial dataset. The binding sites considered, indeed, originate from a selection of the most representative sites carried out in the work of Bovolenta et al.~\cite{bovolenta2022}. Therefore, including pairs with a high  distance would imply direct hops between sites that may, in reality, have in between other unknown states, since not all energetic minima have been explicitly characterized. Finally, in order to limit the computational cost, a representative set of 27 site pairs was selected, chosen to cover a wide variety of geometric and energetic configurations. In particular, the 27 representative pairs were chosen to cover the whole range of inter-site distances (up to 5~\AA) and different binding energy combinations (low–low, low–high, high–high), to make sure that the sampled pathways preserved the nature of the initial data set.
\end{enumerate}
After the network of plausible diffusion pathways is refined through the filtering criteria, it is possible to proceed with a detailed quantum-chemical characterization of the selected transitions.

\subsection{Diffusion rates calculation}
\begin{figure}[htbp]
\centering

\begin{subfigure}[b]{0.40\linewidth}
\includegraphics[width=\linewidth]{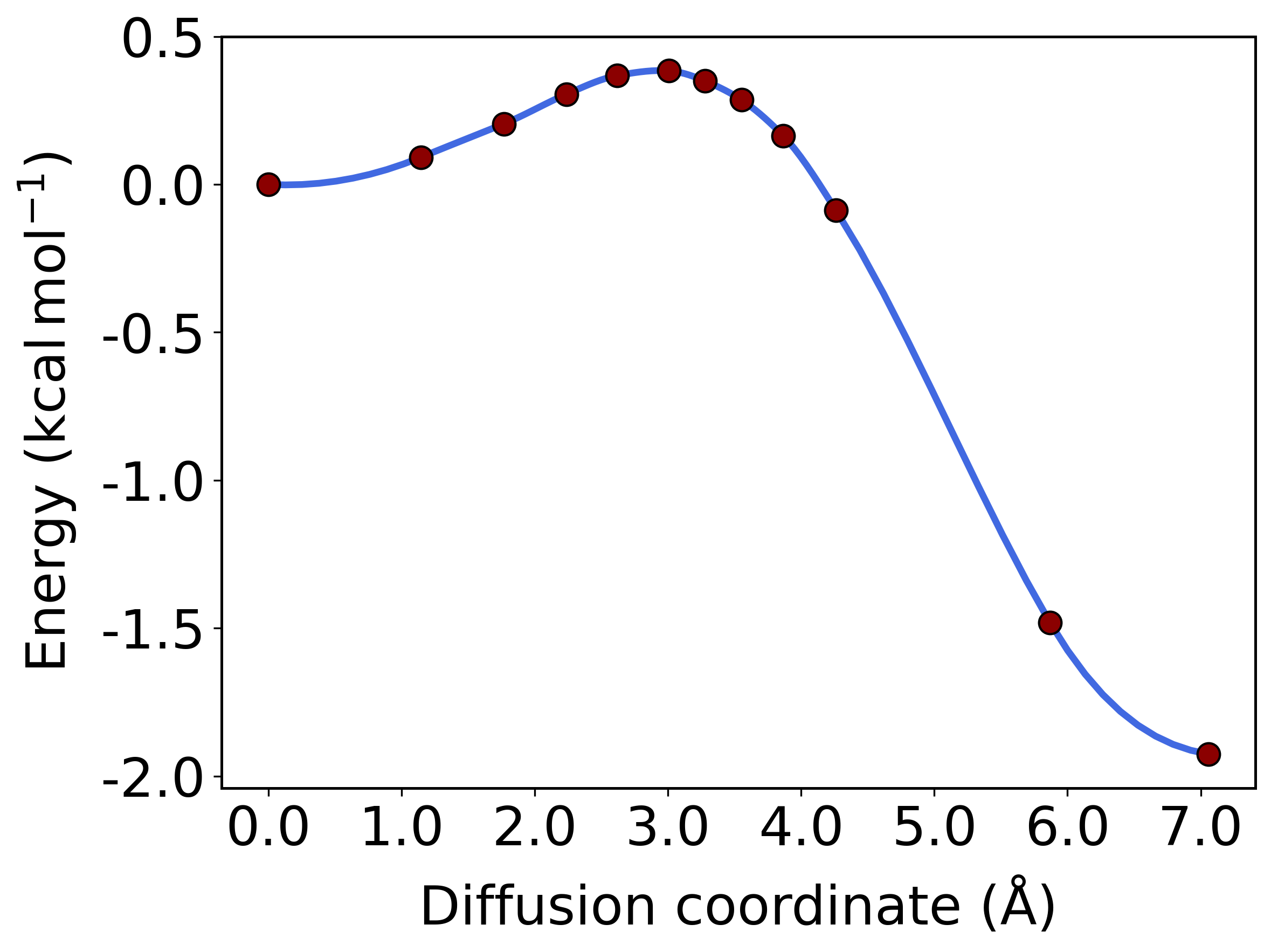}
\caption{}
\label{subfig:mep1}
\end{subfigure}
\hspace{2em}
\begin{subfigure}[b]{0.40\linewidth}
\includegraphics[width=\linewidth]{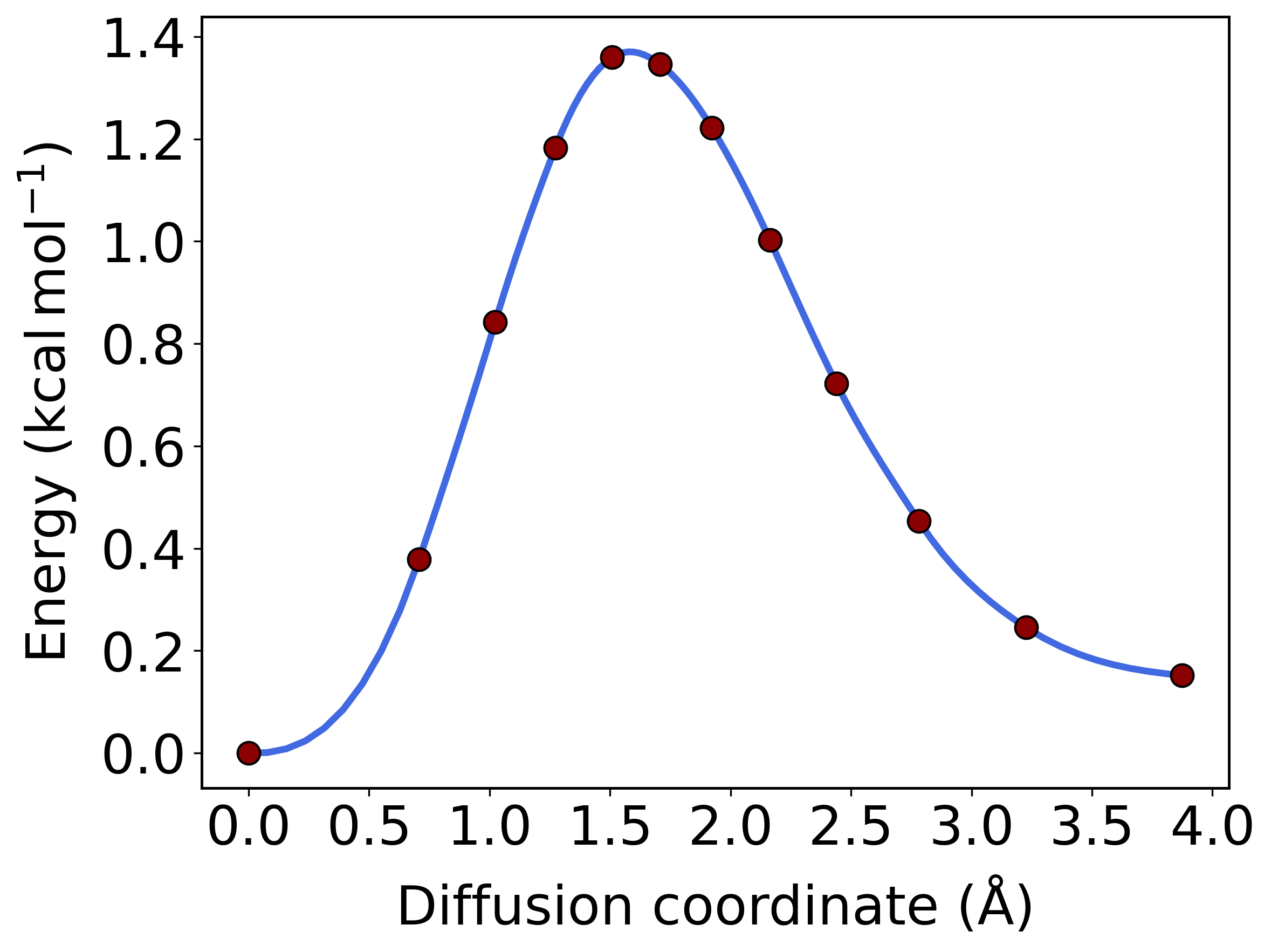}
\caption{}
\label{subfig:mep2}
\end{subfigure}

\vspace{0.5em}

\begin{subfigure}[b]{0.40\linewidth}
\includegraphics[width=\linewidth]{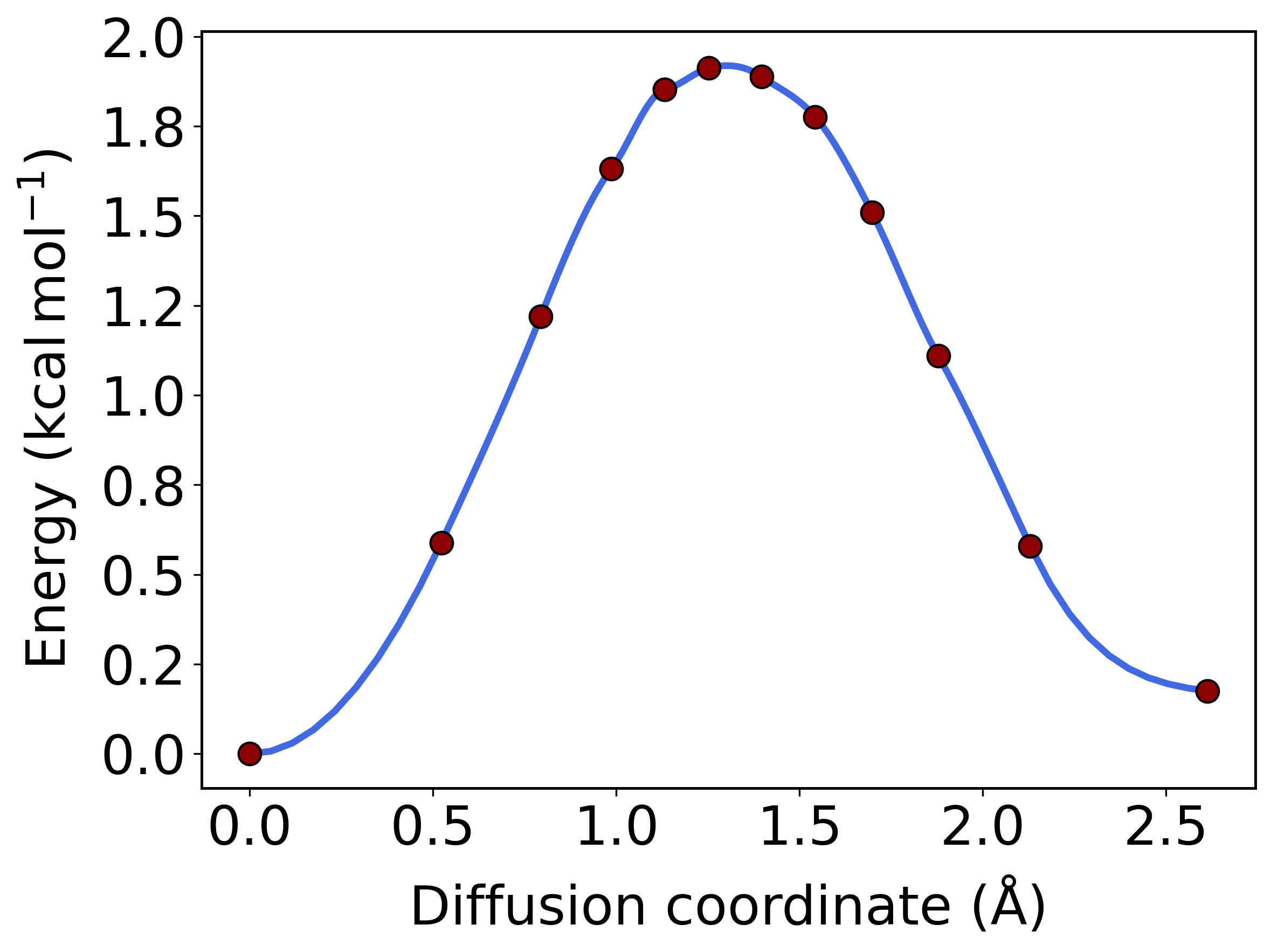}
\caption{}
\label{subfig:mep3}
\end{subfigure}
\hspace{2em}
\begin{subfigure}[b]{0.40\linewidth}
\includegraphics[width=\linewidth]{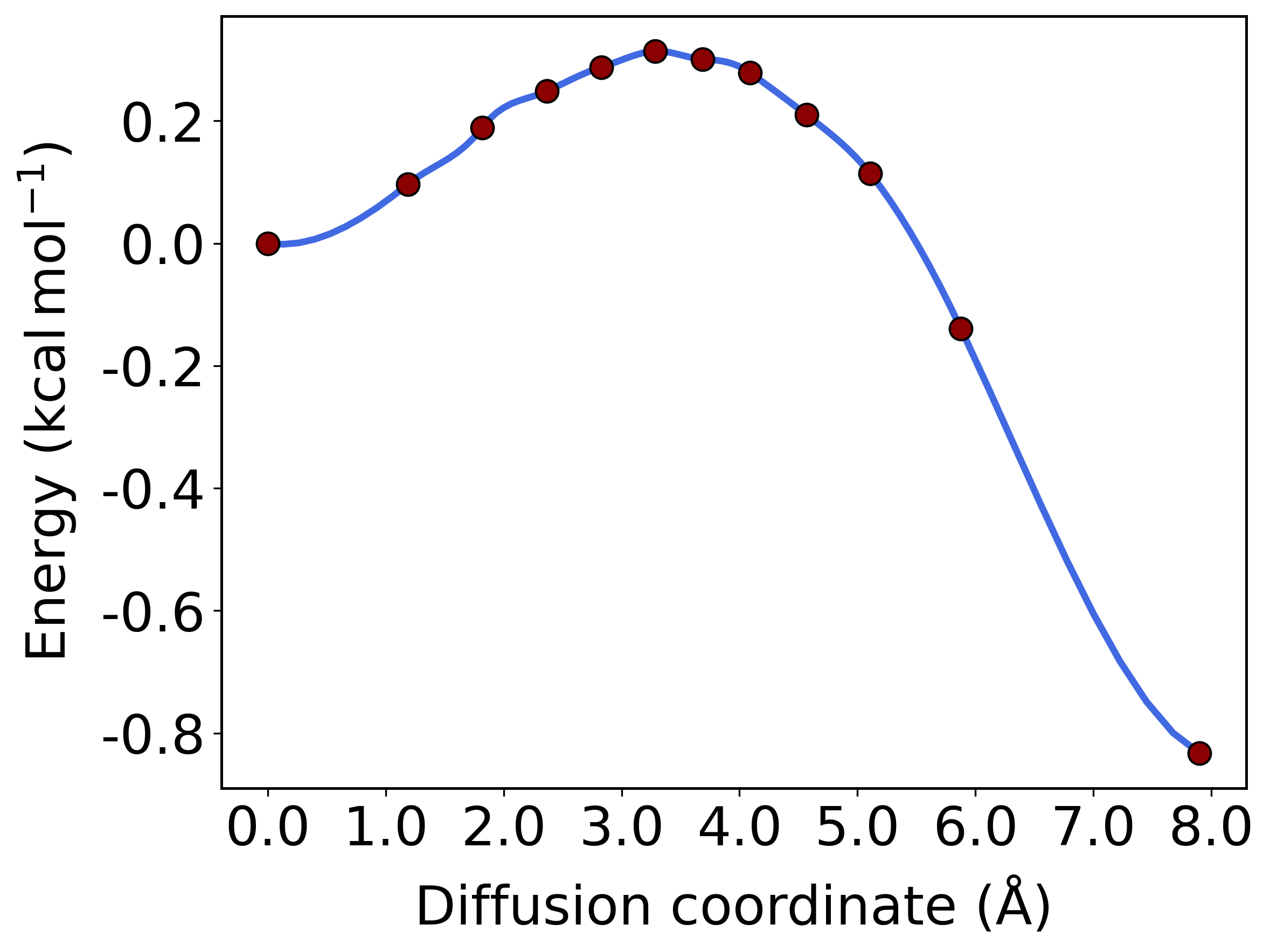}
\caption{}
\label{subfig:mep4}
\end{subfigure}

\caption{\label{fig:meps}Energy profiles computed with the NEB method for 4 representative diffusion events: (\subref{subfig:mep1}), (\subref{subfig:mep2}), (\subref{subfig:mep3}) and (\subref{subfig:mep4}). In each plot, the red dots represent the energies of the 12 images (including the initial and final states), while the blue line indicates the MEP interpolation. Each panel uses an independent energy scale for better visualization.}
\end{figure}

\begin{figure}[htbp]
\centering

\makebox[0.6\linewidth][c]{%
  \hspace{-0.01\linewidth}\textbf{Initial site}%
  \hspace{0.13\linewidth}\textbf{TS}%
  \hspace{0.15\linewidth}\textbf{Final site}%
}

\noindent\rule{0.6\linewidth}{0.2pt}

\begin{subfigure}[b]{0.6\linewidth}
\includegraphics[width=\linewidth]{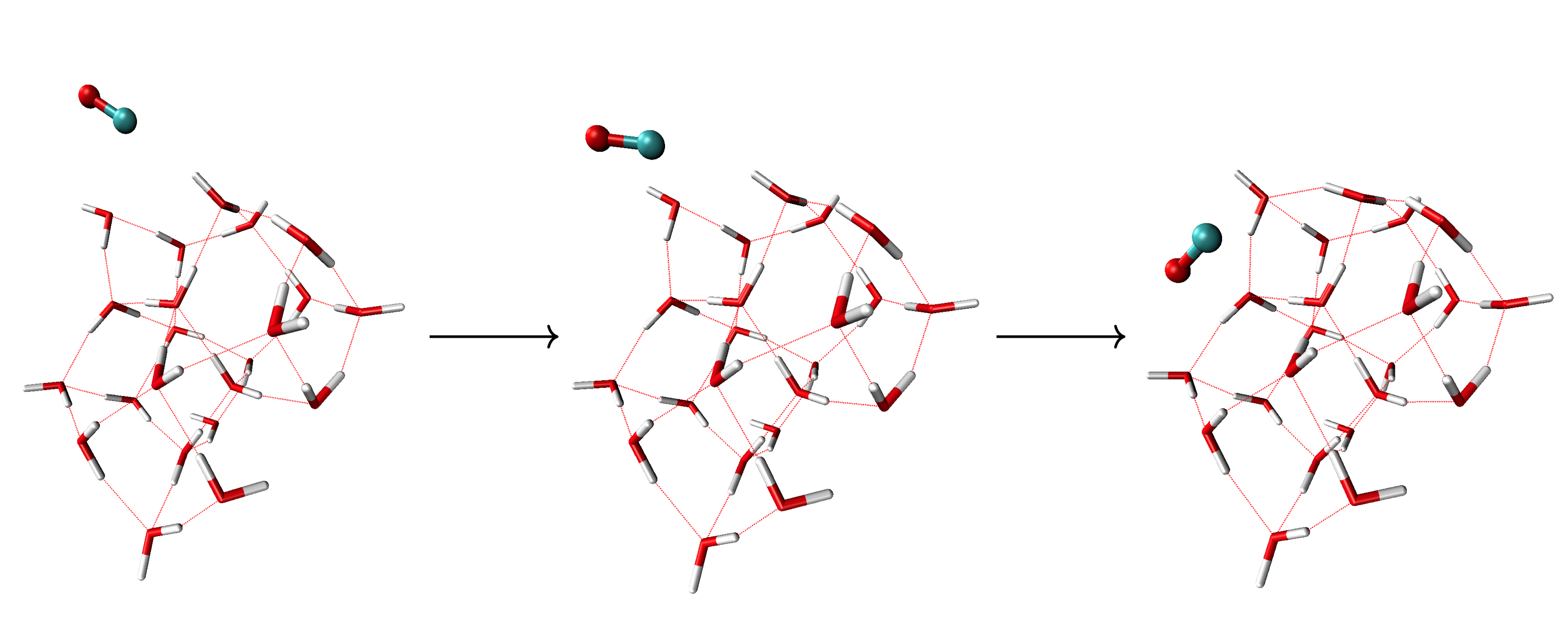}
\caption{}
\label{subfig:12_31}
\end{subfigure}

\begin{subfigure}[b]{0.6\linewidth}
\includegraphics[width=\linewidth]{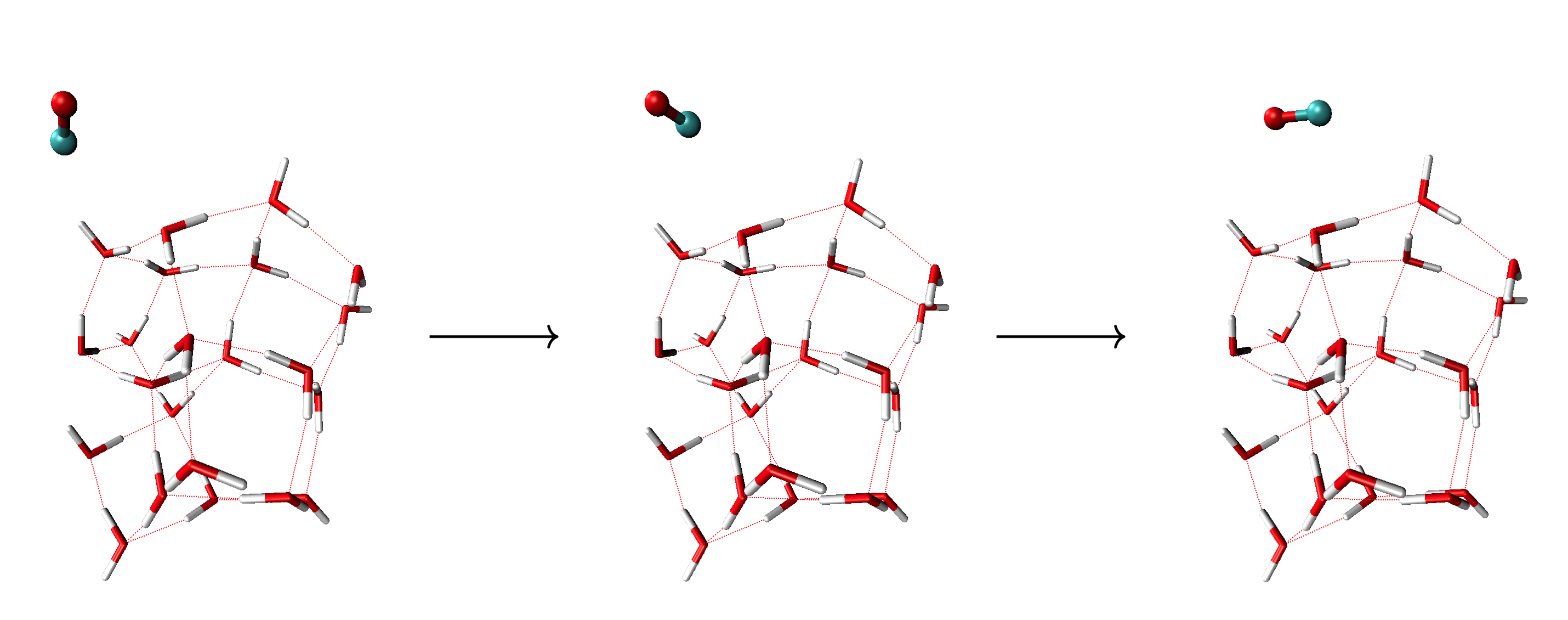}
\caption{}
\label{subfig:18_06}
\end{subfigure}

\begin{subfigure}[b]{0.6\linewidth}
\includegraphics[width=\linewidth]{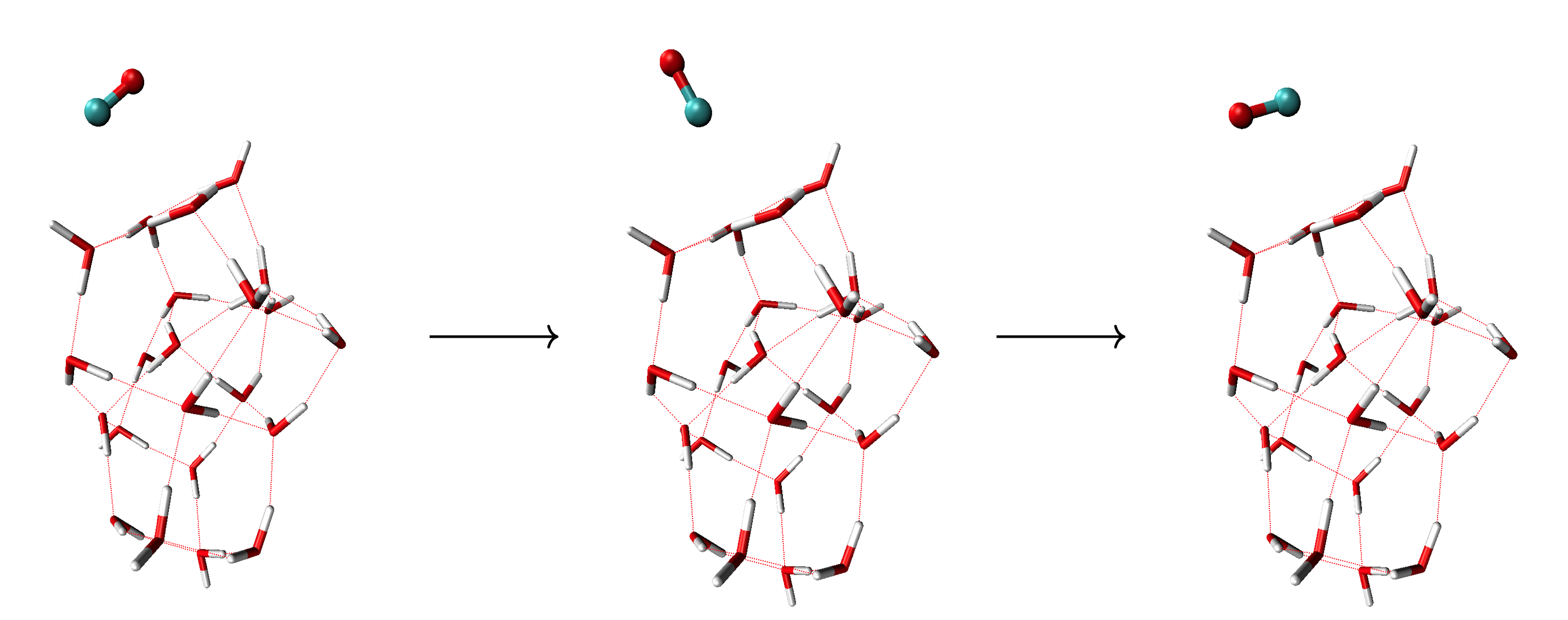}
\caption{}
\label{subfig:31_75}
\end{subfigure}

\begin{subfigure}[b]{0.6\linewidth}
\includegraphics[width=\linewidth]{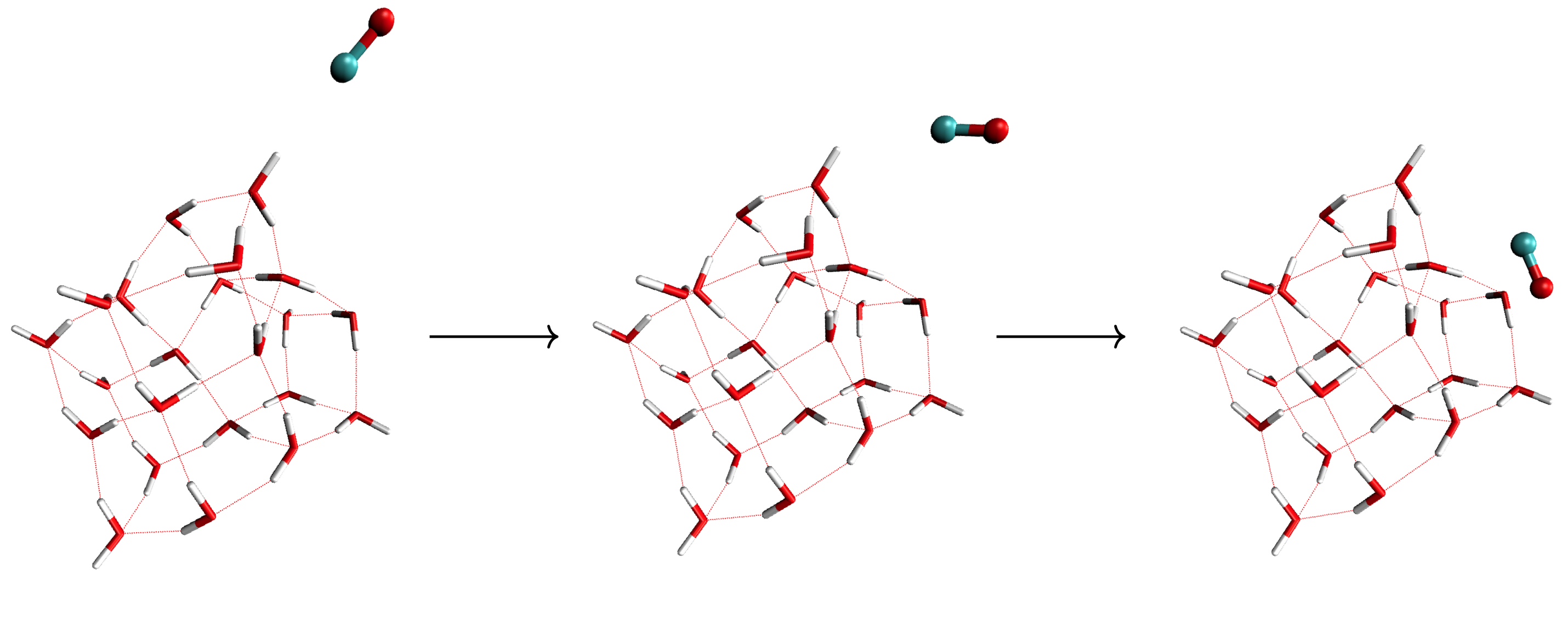}
\caption{}
\label{subfig:13_09}
\end{subfigure}

\caption{\label{fig:clusters}Optimized structures (computed at M062X-D3/def2-TZVP level) of the initial and final binding sites and the corresponding transition state for the 4 representative diffusion events shown in Fig.~\ref{fig:meps}: (\subref{subfig:12_31}), (\subref{subfig:18_06}), (\subref{subfig:31_75}) and (\subref{subfig:13_09}). Each subfigure shows the initial minimum (left), the transition state (center), and the final minimum (right).}
\end{figure}

\begin{figure}[htbp]
    \center
    \includegraphics[scale=0.45]{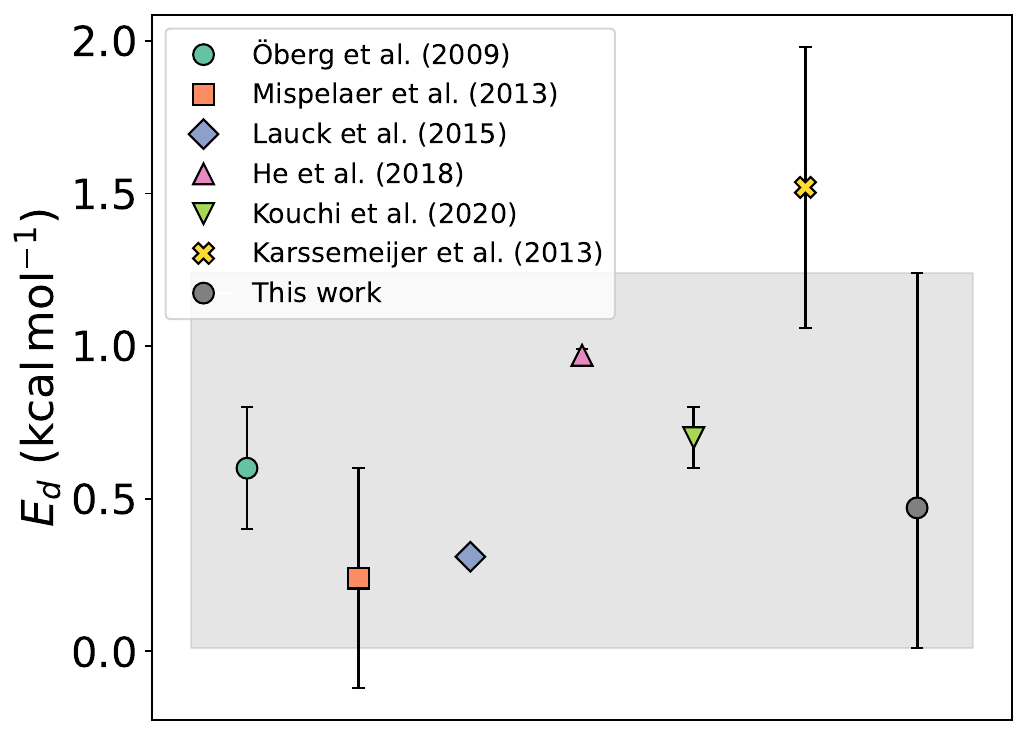}
    \caption{\label{fig:ed_comparison}Box plot comparison between the diffusion energy barriers for CO on ASW presented in this work and the values reported in previous investigations. The gray shaded area indicates the range of values obtained for $E_\mathrm d$ in this study.}
\end{figure}

Fig.~\ref{fig:meps} shows four representative examples of energy profiles obtained from the NEB calculations, corresponding to diffusion events happening on different ASW clusters, chosen to highlight the energetic and geometrical diversity of the selected transitions. For the same site pairs, Fig.~\ref{fig:clusters} reports the optimized geometries of the initial and final binding sites together with the corresponding transition states.
The energy barriers, obtained at MPWB1K-D3BJ/def2-TZVP level, for the different pairs of sites are reported in Tab.~\ref{barriers}, with values ranging from almost zero up to 1.24~kcal\,mol$^{-1}$ (624~K), and an average of 0.47~kcal\,mol$^{-1}$ (237~K).
Additionally, a comparison with the values obtained from the previous studies is presented in Fig.~\ref{fig:ed_comparison}, showing good agreement with available experimental estimates.

\begin{table}[htbp]
\centering
\caption{\label{barriers}Diffusion energy barriers, computed at MPWB1K-D3BJ/def2-TZVP level, for the selected pairs of sites of ASW clusters.
For each calculation, their distance (in \AA), and the forward and backward activation barriers (in kcal\,mol$^{-1}$) are reported.
Missing values indicate that the estimated barrier is below the accuracy threshold of the adopted computational method.
The data are presented in two subtables, grouping distances below 4~\AA~(a) and distances between 4 and 5~\AA~(b).}
\begin{subtable}[t]{0.45\textwidth}
  \subcaption{\label{subtable_1}}
  \centering
  \begin{tabular}{ccc}
    \hline
    $d$ & $\Delta E^\ddagger_{\text{dir}}$ & $\Delta E^\ddagger_{\text{inv}}$ \\
    \hline
    2.24  & 0.04 & 0.03 \\
    2.26 & - & 0.12 \\
    2.31 & 0.59 & 0.66 \\
    2.57 & 0.65 & 1.03 \\
    2.72 & 0.28 & - \\
    3.28 & 0.65 & 0.86 \\
    3.42 & 0.01 & 0.02 \\
    3.45 & 0.50 & 0.02 \\
    3.51 & 0.24 & - \\
    3.64 & 0.18 & 0.17 \\
    3.65 & 0.18 & 0.01 \\
    3.71 & 0.11 & 0.19 \\
    3.80 & 0.58 & 0.68 \\
    3.87 & 0.63 & 0.83 \\
    \hline
  \end{tabular}
\end{subtable}
\quad
\begin{subtable}[t]{0.45\textwidth}
  \subcaption{\label{subtable_2}}
  \centering
  \begin{tabular}{ccc}
    \hline
    $d$ & $\Delta E^\ddagger_{\text{dir}}$ & $\Delta E^\ddagger_{\text{inv}}$ \\
    \hline
    4.04 & 0.71 & 0.84 \\
    4.13 & 0.90 & 1.06 \\
    4.16 & - & 0.60 \\
    4.21 & - & 0.72 \\
    4.26 & 0.10 & 0.03 \\
    4.32 & 0.34 & 0.38 \\
    4.34 & - & 0.21 \\
    4.43 & 0.61 & 0.89 \\
    4.54 & 0.06 & 0.68 \\
    4.55 & - & 0.53 \\
    4.58 & 0.53 & 1.24 \\
    4.67 & - & 0.90 \\
    4.73 & 0.23 & 0.78 \\
    \hline
  \end{tabular}
\end{subtable}
\end{table}

\begin{figure}[ht]
\centering
\begin{subfigure}[b]{0.45\linewidth}
\includegraphics[width=\linewidth]{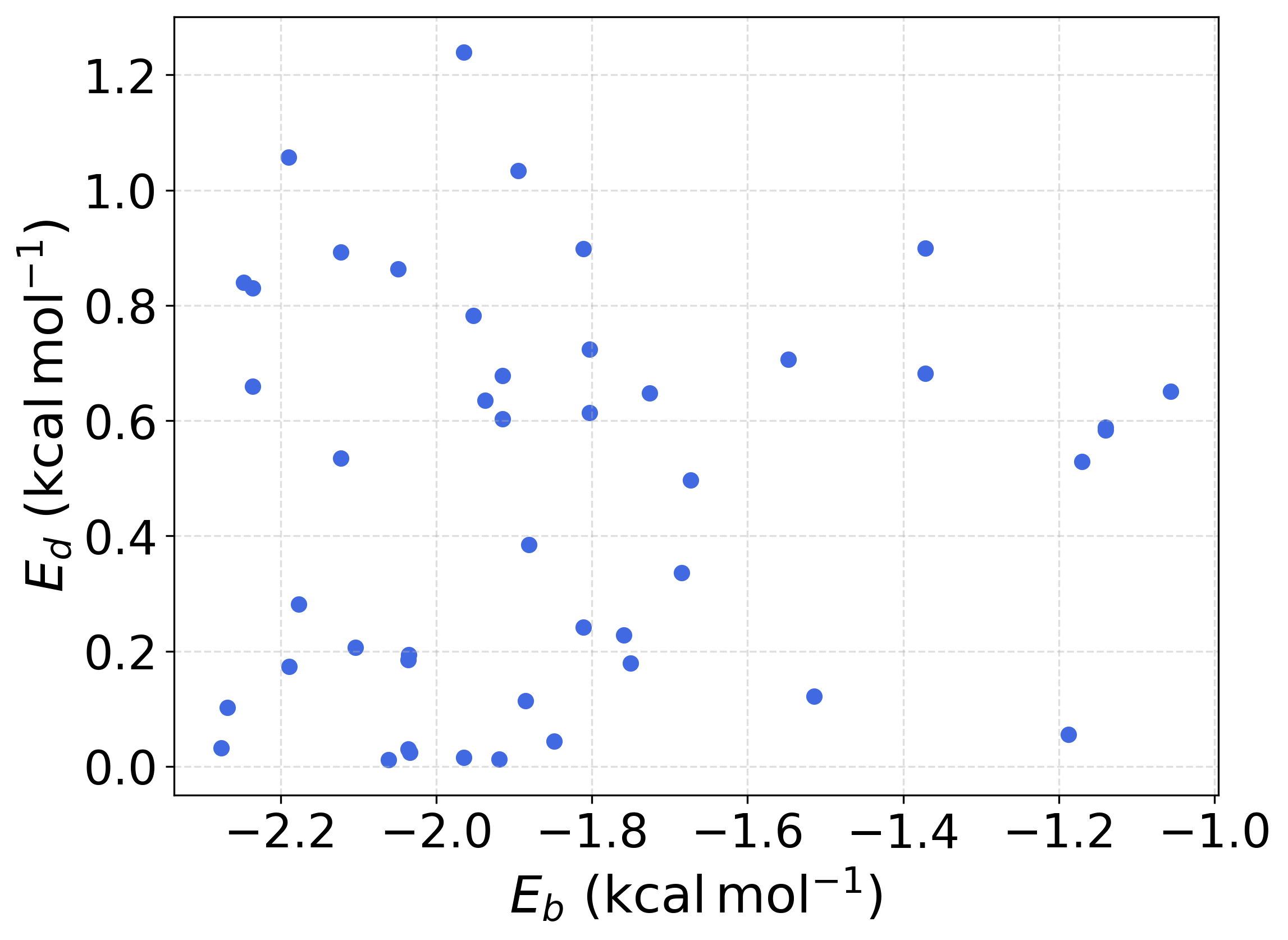}
\caption{}
\label{subfig:ed_vs_eb}
\end{subfigure}
\hspace{2em}
\begin{subfigure}[b]{0.44\linewidth}
\includegraphics[width=\linewidth]{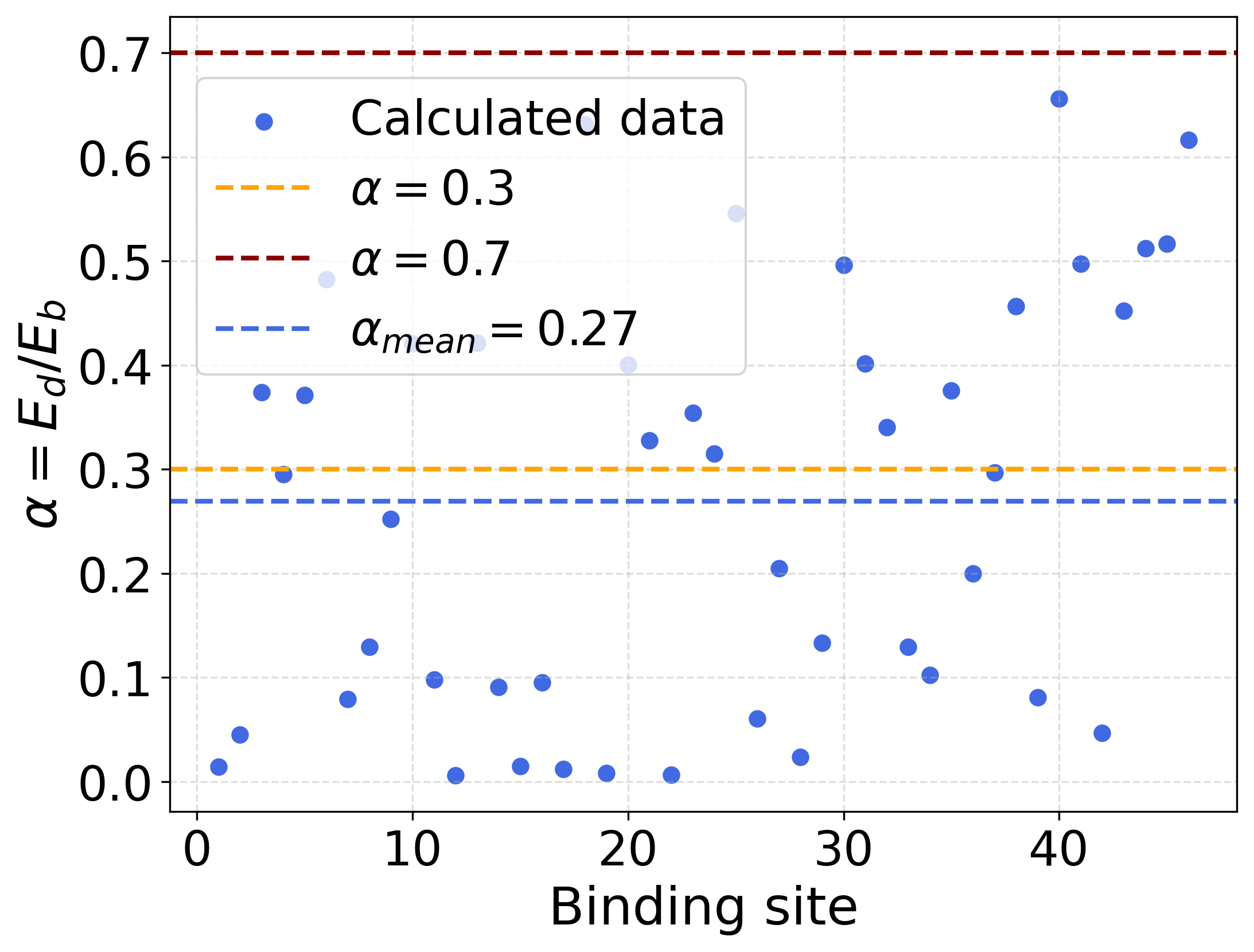}
\caption{}
\label{subfig:alpha_plot}
\end{subfigure}
\caption{(\subref{subfig:ed_vs_eb}) Distribution of the diffusion energy ($E_{\mathrm{d}}$) as a function of the binding energy ($E_{\mathrm{b}}$) for the selected pairs of adsorption sites. (\subref{subfig:alpha_plot}) Calculated values of $\alpha$ for the selected pairs of adsorption sites. Along with the obtained mean value, the two lines corresponding to the commonly assumed limiting values of $\alpha$ (0.3 and 0.7) are also plotted, and it is evident how the data is not strictly confined between the two limits. See further details in the main text.}
\label{fig:ed_vs_eb}
\end{figure}

A relevant insight emerges from the comparison between the diffusion energies ($E_\mathrm{d}$) and the binding energies ($E_\mathrm{b}$) of the initial adsorption sites, shown in Fig.~\ref{subfig:ed_vs_eb}. Contrary to what is commonly assumed in astrochemical models, the data obtained from the DFT calculations do not highlight a linear correlation between these two parameters. We note that the binding energy distribution employed in this work does not extend to the low-coverage limit measured experimentally, around 1600 K (3.18 kcal$\,\text{mol}^{-1}$)~\cite{he2016binding}. This is because only a very small fraction of the binding sites sampled by Bovolenta et al.~\cite{bovolenta2022} fall within the high-energy tail of the distribution, making these configurations statistically rare compared to the dominant population around 1035 K (2.06 kcal$\,\text{mol}^{-1}$).
Thus, the binding energy range shown in Fig.~\ref{subfig:ed_vs_eb} reflects the most probable adsorption environments sampled by CO on ASW rather than the low-coverage limit.
\\
Additionally, in Fig~\ref{subfig:alpha_plot} are reported the $\alpha$ ratios computed for every pair of binding sites, obtained performing the ratio between the diffusion energy barrier and the binding energy of the initial adsorption site.
The commonly used approximation, according to which the ratio $\alpha = E_\mathrm{d} / E_\mathrm{b}$ is between 0.3 and 0.7, corresponding to the two orange and red lines of Fig~\ref{subfig:alpha_plot}, does not appear to be universally valid to describe the diffusion of CO on ASW surfaces. Although the line corresponding to the lower bound ($\alpha = 0.3$) seems to reasonably capture the average value of the dataset, it is important to point out that even small deviations in the diffusion energy barriers can propagate into very large errors for the rate coefficients, due to the exponential dependence of Eq.~\ref{diff_arrhenius}. Therefore, the commonly adopted approximation does not seem to adequately represent the physico-chemical properties of interstellar ices.

Additionally, from the DFT calculations presented above, the mean $\alpha$ ratio obtained for CO molecules is 0.27. This value, being significantly lower than those commonly employed in kinetic models, suggests that the contribution of diffusion in surface-mediated reactivity could be more important than previously assumed.

\subsection{Comparison with desorption rates}
\begin{figure}[ht]
\centering
\begin{subfigure}[b]{0.45\linewidth}
\includegraphics[width=\linewidth]{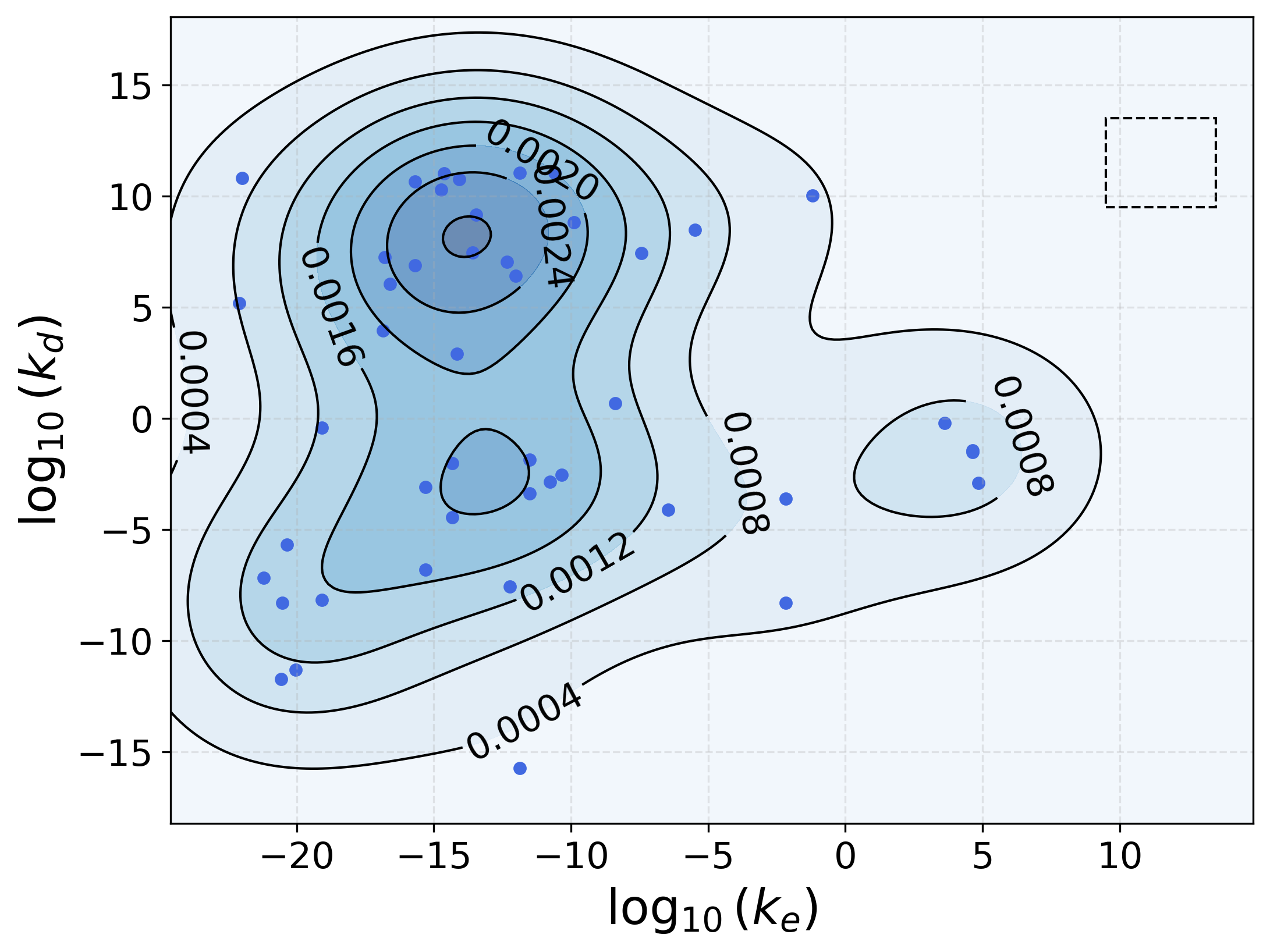}
\caption{}
\label{subfig:kd_vs_ke_10K}
\end{subfigure}
\hspace{2em}
\begin{subfigure}[b]{0.45\linewidth}
\includegraphics[width=\linewidth]{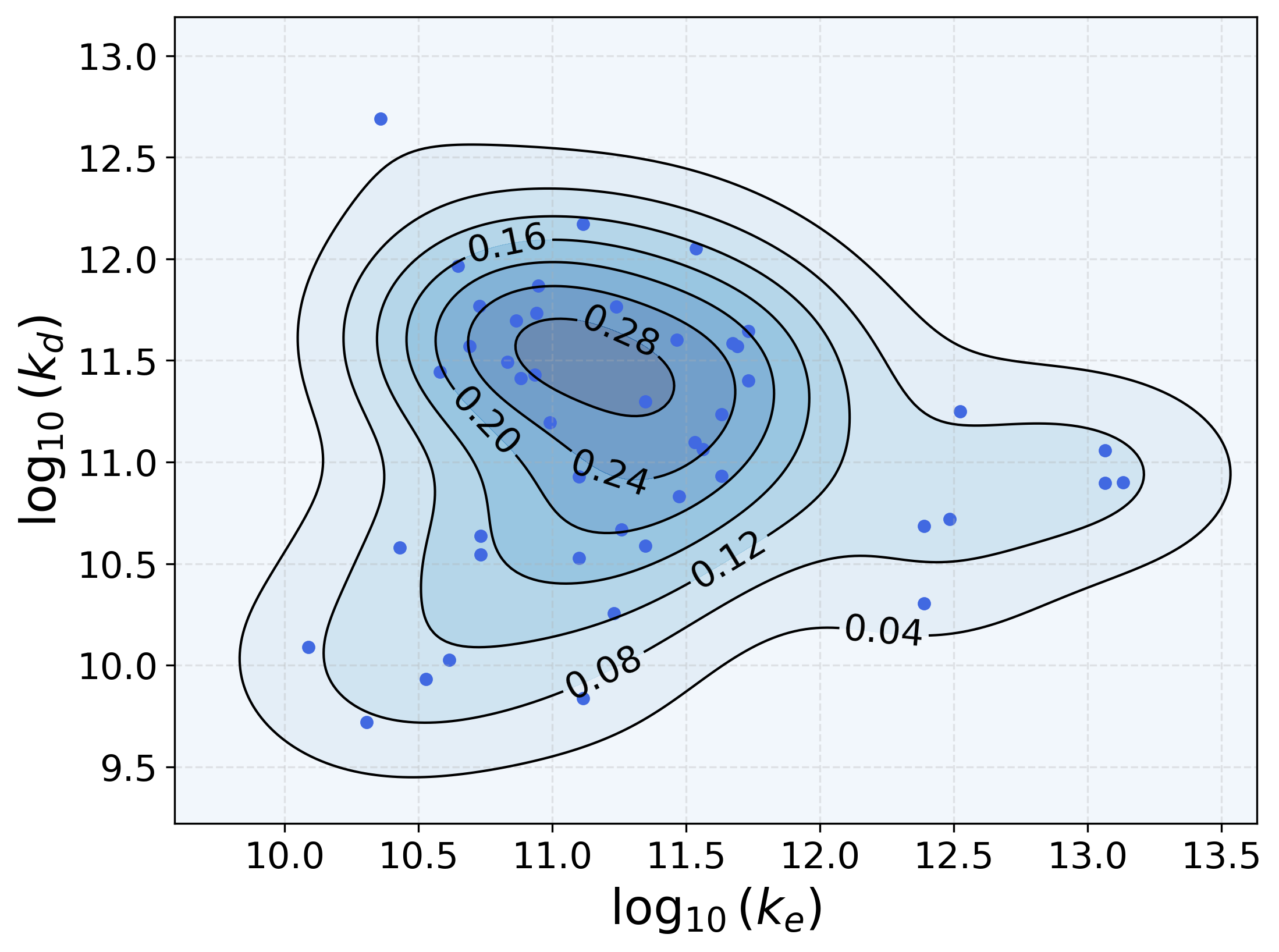}
\caption{}
\label{subfig:kd_vs_ke_100K}
\end{subfigure}
\caption{Scatter plot between diffusion ($k_\mathrm d$) and desorption ($k_\mathrm e$) rate coefficients for the different pairs of binding sites at 10~K (\subref{subfig:kd_vs_ke_10K}) and 100~K (\subref{subfig:kd_vs_ke_100K}). For each plot, the KDE performed on the point distribution is shown as shaded contours. The dashed box in the left-side panel spans the region covered by the distribution at 100~K, highlighting the changes induced by the temperature variation.}
\label{fig:kd_vs_ke}
\end{figure}

Calculating diffusion and desorption rate coefficients (through Eq.~\ref{k_diff} and Eq.~\ref{des_rate}) allows a direct comparison between the two competing processes, which gives key insights into the equilibrium between surface mobility and retention under interstellar conditions.
Moreover, temperature plays a crucial role in regulating this equilibrium, and its influence can be better established by examining the scatter plots between the two rate coefficients, shown in Fig.~\ref{fig:kd_vs_ke} for two representative temperatures, $T=10$~K and $T=100$~K, respectively.

It is clear how, at low temperatures, diffusion appears to be systematically faster than desorption and, by increasing the temperature, the two processes start to happen on similar time scales. The main reason behind this trend lies in the difference between the diffusion barrier and the binding energy for each site. Increasing the temperature, the exponential term in Eq.~\ref{k_diff} and Eq.~\ref{des_rate} becomes less dominant, and, as a consequence, the difference between $k_\mathrm{d}$ and $k_\mathrm{e}$ decreases. 
To obtain a continuous representation of the density probability of the rates from the discrete data set, a bidimensional Kernel Density Estimation (KDE) was performed on the ($k_\mathrm{d},k_\mathrm{e}$) distributions, and the results are also shown in Fig.~\ref{fig:kd_vs_ke}. In this context, obtaining a joint probability distribution function of $(k_\mathrm{d},k_\mathrm{e})$ is crucial to provide a more realistic description of surface dynamics, especially in astrophysical environments where the temperature changes smoothly in time and space (such as star- and planet-forming regions), affecting the relative probabilities of diffusion and desorption.
It is possible to notice how, increasing the temperature, the  distribution moves towards higher values of diffusion and desorption rate coefficients, according to faster kinetics. Moreover, the distribution tends to become narrower, and this reduced dispersion is due to the weaker influence of the exponential term at higher temperatures. 
However, while this analysis provides a first qualitative insight into the temperature dependence of diffusion and desorption, the use of these data in astrochemical modeling requires a continuous statistical description of these rate distributions, which is addressed in the next Section.

\section{Astrophysical implications}
\label{sec_astro}
In order to explore the effect of the calculated kinetic parameters in an idealized case, we applied the computed values to a simple one-dimensional model. 
At first, the KDE distributions have been approximated as 2D Gaussian functions at different temperatures. The trends of the Gaussian parameters (amplitude, center coordinates, and standard deviations) as functions of temperature were interpolated using polynomial functions, allowing their integration into the numerical code. The evaluation of the Gaussian fitting accuracy is discussed in Appendix~\ref{app_C}.

\begin{figure}[htbp]
    \centering
    \includegraphics[scale=0.50]{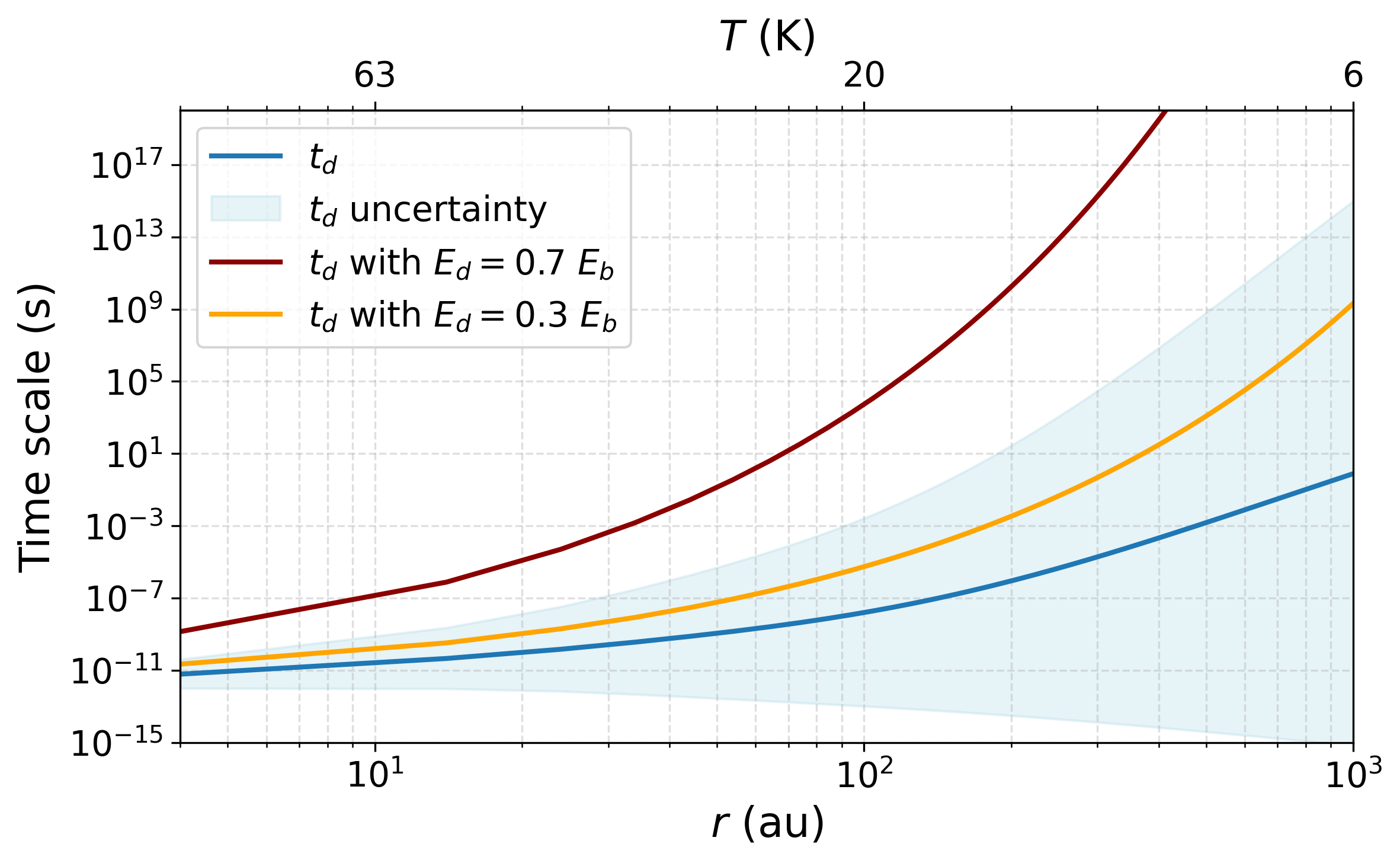}
    \caption{Surface diffusion time-scales for CO as a function of the distance from the central star. The blue curve represents the values obtained from the centroid of the computed Gaussian distributions, while the red and orange curves represent the diffusion time-scales calculated assuming fixed $\alpha$ ratios of 0.7 and 0.3, respectively. The shaded blue area represents the width of the rate coefficients' distribution at each temperature.}
    \label{fig:timescales_diff}
\end{figure}

Starting from these probability distributions, the time-scales for the diffusion process were computed at different temperatures. In particular, to evaluate the impact of the newly computed parameters in an astrophysically relevant scenario, the midplane of a static protoplanetary disk was taken into consideration,
where the temperature of the dust decreases with the distance
from the center of the star~\cite{oberg2023protoplanetary}.
\\
For each value of $r$, the diffusion time-scales were computed as the inverse of the respective rate coefficients, assuming a temperature radial profile typical of the Minimum Mass Solar Nebula (MMSN) model (see Appendix~\ref{app_D} for more details). In particular, the rate coefficients corresponding to the centroid of the Gaussian distribution have been compared to the values commonly assumed in astrochemical models. This comparison is shown in Fig.~\ref{fig:timescales_diff}, where the diffusion time-scales derived from the Gaussian centroids result in significantly lower values than the ones obtained assuming the often adopted ratio $\alpha=0.7$~\cite{semenov2010chemistry,ruffle2000new}.

This result highlights that the diffusion energy barriers obtained in this work are systematically lower than those commonly assumed in astrochemical models, suggesting that the widely used approximation may underestimate the surface mobility of molecules under interstellar conditions. This insight has a series of implications that range from the description of surface reactivity to the prediction of structural parameters for astrophysical objects. Firstly, the magnitude of diffusion energy barriers strongly affects surface reactivity, since the overall reaction rate depends on the mobility of the two reactants~\cite{semenov2010chemistry}. Thus, an enhanced diffusion dynamics could become particularly significant when considering the surface-mediated synthesis of interstellar COMs.
\\
Moreover, an increase in surface mobility of the species has an impact on the desorption dynamics. Assuming a Gaussian distribution for the binding energies, at a given dust thermal energy $E_{\mathrm{thermal}} ^{\mathrm{dust}}$\,, species located in the binding sites with $E_\mathrm b < E_{\mathrm{thermal}} ^{\mathrm{dust}}$ (low-energy tail of the distribution) will undergo a fast thermal desorption process. The species located at binding sites with $E_\mathrm b > E_{\mathrm{thermal}} ^{\mathrm{dust}}$ (high-energy tail of the distribution), instead, will tend to remain adsorbed on the surface, lacking the necessary thermal energy to efficiently desorb. However, the model introduced by Grassi et al.~\cite{grassi2020} that tracks a binding-energy distribution can be extended by introducing the contribution of surface mobility.
After the initial desorption from the low-energy tail of the distribution, surface diffusion might allow molecules from the high-energy sites to migrate to newly available lower-energy sites. Once relocalized, these molecules might assume a new configuration that allows them to desorb, even if they originally belonged to a strongly-bound site. Thus, surface diffusion could substantially affect the desorption dynamics, enhancing the overall efficiency of the process.
\\

As an observable consequence, this enhancement in desorption efficiency may significantly affect the position of snowlines in protoplanetary disks, defined as the region where a given species undergoes the transition from solid to gas~\cite{oberg2021astrochemistry}. Surface diffusion enables molecules to migrate toward sites with lower binding energy, promoting desorption even at relatively low-temperature conditions. As a consequence, the inclusion of diffusion processes is expected to shift the CO snowline outward, moving the gas–solid transition to larger radial distances. Incorporating surface diffusion explicitly into disk models would thus be particularly valuable, as it would improve the accuracy of snowline predictions and provide deeper insight into the coupling between grain-surface chemistry and the dynamical evolution of protoplanetary disks. While an enhanced diffusion is expected to shift the CO snowline outward, a quantitative estimate of this effect would require a multi-binding/multi-diffusion disk modeling approach, which is beyond the scope of this paper and is left for future developments.
Finally, a more efficient diffusion-induced desorption, will decrease the amount of molecules on the surface, with a catastrophic effect on the overall reactivity and formation of more complex species on surface.

\section{Conclusions}
\label{sec_conclusions}
In the present work, we have investigated by  quantum-chemical calculations CO's surface diffusion processes on ASW clusters, representative of interstellar ices. The employed computational pipeline allowed the accurate estimate of the energy barriers associated with diffusion hops, highlighting a significant heterogeneity with values ranging from almost zero up to 1.24 kcal\,mol$^{-1}$ (624~K), and an average of 0.47 kcal\,mol$^{-1}$ (237~K). This average value is significantly lower than the assumptions typically adopted in astrochemical models, where diffusion energies are usually approximated as a constant fraction of the binding energy.
\\
The comparison with the few available experimental data shows a good agreement with the lowest estimates reported in the literature, supporting the reliability of the adopted computational approach. The obtained results indicate how CO mobility on ASW surfaces could be much higher than previously assumed, with thermally-activated diffusion events already effective at typical temperatures of molecular clouds (10–20~K). This suggests a potentially significant impact on the formation of COMs as well as on the dynamics of desorption processes.
\\
Surface diffusion represents a crucial process for the determination of molecular abundances in the interstellar medium. Adopting more realistic values for its kinetic parameters in astrochemical models will substantially reduce the current uncertainties, thus improving their predictive capabilities. A possible extension of this approach to different molecules, in particular radical species, will help to further clarify the role of surface mobility in the formation of COMs and, more broadly, in the chemical evolution of star- and planet-forming regions.

\newpage

\begin{acknowledgement}
FB acknowledges financial support from the Faculty of Science of Sapienza University of Rome and is thankful to Paola Caselli for having hosted him at the Center for Astrochemical Studies, MPE, where part of this work has been pursued.
SB acknowledges BASAL Centro de Astrofisica y Tecnologias Afines (CATA), project number AFB-17002.
\end{acknowledgement}

\section*{Conflict-of-interest statement}
The authors declare no competing financial interests.

\begin{suppinfo}
We provide the \verb|.xyz| coordinates for all the stationary points reported in this work. For the energy minima, the file are named as \verb|ClusterIndex_SiteIndex.xyz|, while for the saddle points, they are named as \verb|ClusterIndex_InitialSiteIndex_FinalSiteIndex.xyz|.
\end{suppinfo}

\bibliography{achemso-demo}

\newpage

\begin{appendices}

\section{Calculation of partition functions}
\label{app_A}
The calculation of diffusion and desorption rate coefficients through hTST requires the evaluation of the roto-vibrational partition functions for initial, transition, and final states. In this work, they are computed using custom \textsc{Python} scripts, and the details on the employed formulas are provided in this section.

For a set of vibrational frequencies $\{\nu_i\}$, in cm$^{-1}$, the vibrational partition function $q_{\mathrm{vib}}$ is given by
\begin{equation}
    q_{\mathrm{vib}} = \prod_{j=1}^{3N-6}{\frac{1}{1- e^{-\beta \nu_j}}} \,,
\end{equation}
where $N$ is the number of atoms and
\begin{equation}
    \beta = \frac{hc}{k_\mathrm B T} \,,
\end{equation}
with $h$ being the Planck constant and $c$ the speed of light.

The rotational partition functions $q_\mathrm{rot}$ are computed, within the rigid rotor approximation, as
\begin{equation}
    q_{\mathrm{rot}} = \frac{\sqrt{\pi}}{\sigma} \frac{T^{3/2}}{\sqrt{\theta_A \theta_B \theta_C}} \,,
\end{equation}
where $\sigma$ is the rotational symmetry number and the generic rotational temperature $\theta_r$ is computed as
\begin{equation}
    \theta_r = \frac{hcB}{k_\mathrm B} \,,
\end{equation}
with $B$ being the rotational constant in cm$^{-1}$.

Subsequently, diffusion and desorption rate coefficients are determined as a function of temperature by applying Eq.~\ref{k_diff} and Eq.~\ref{des_rate}, where the total (roto-vibrational) partition function is given by $Q= q_\mathrm{vib} \cdot q_\mathrm{rot}$.

\section{Rate temperature dependence}
\label{app_B}
\begin{figure}[htbp]
\centering

\begin{subfigure}[b]{0.46\linewidth}
\includegraphics[width=\linewidth]{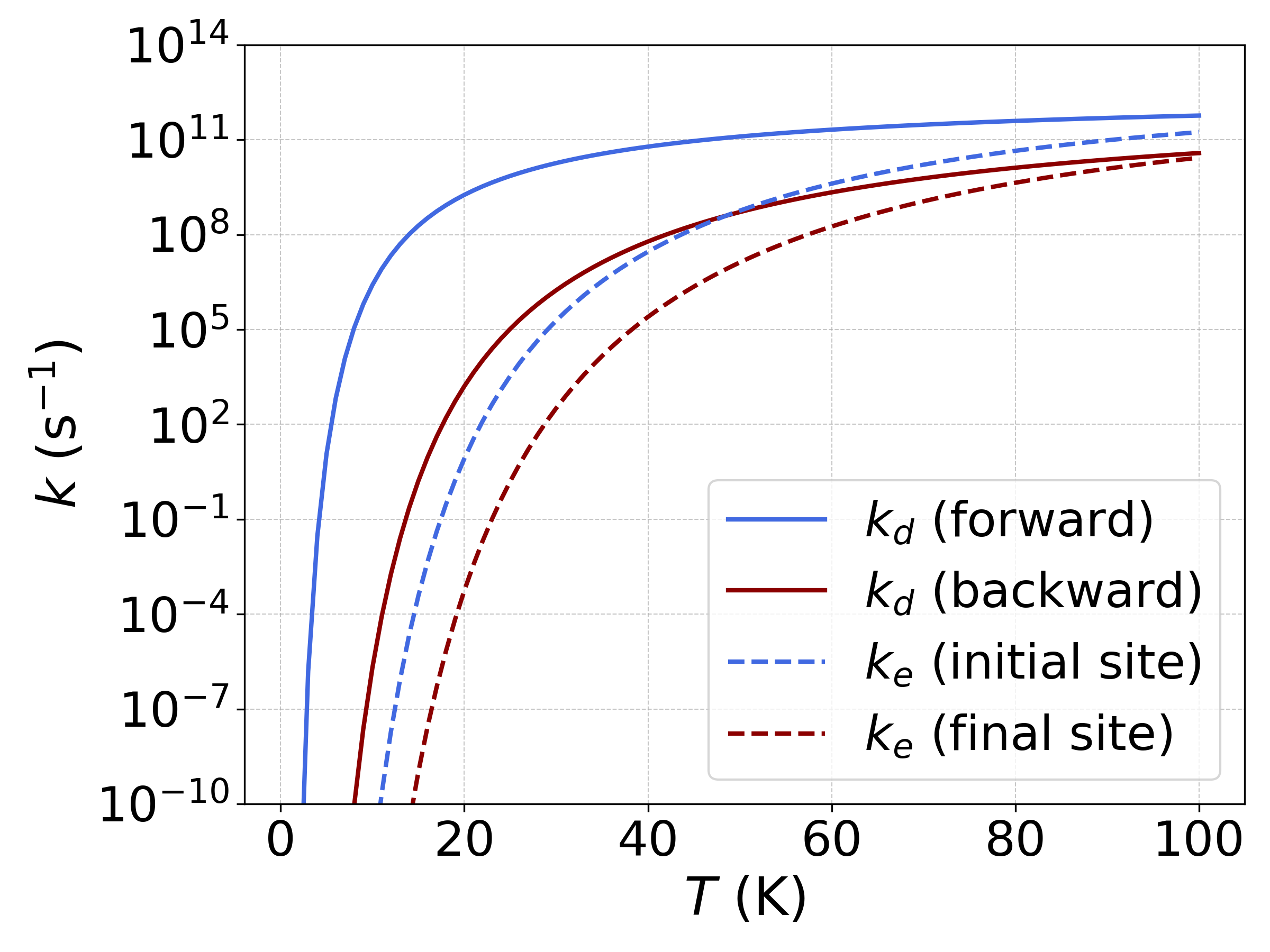}
\caption{}
\label{subfig:diff_vs_des_1}
\end{subfigure}
\hfill
\begin{subfigure}[b]{0.46\linewidth}
\includegraphics[width=\linewidth]{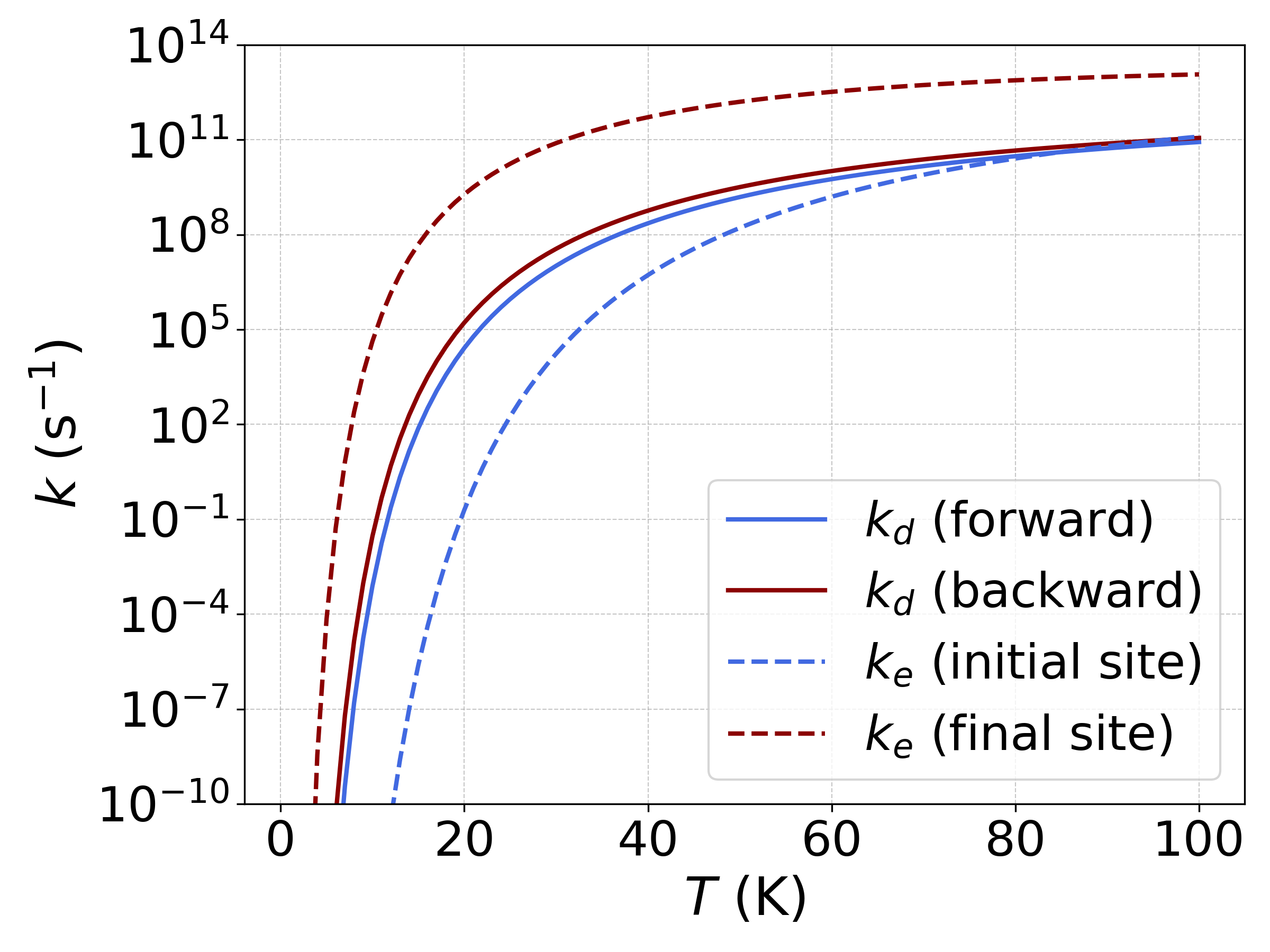}
\caption{}
\label{subfig:diff_vs_des_2}
\end{subfigure}

\vspace{0.5em}

\begin{subfigure}[b]{0.46\linewidth}
\includegraphics[width=\linewidth]{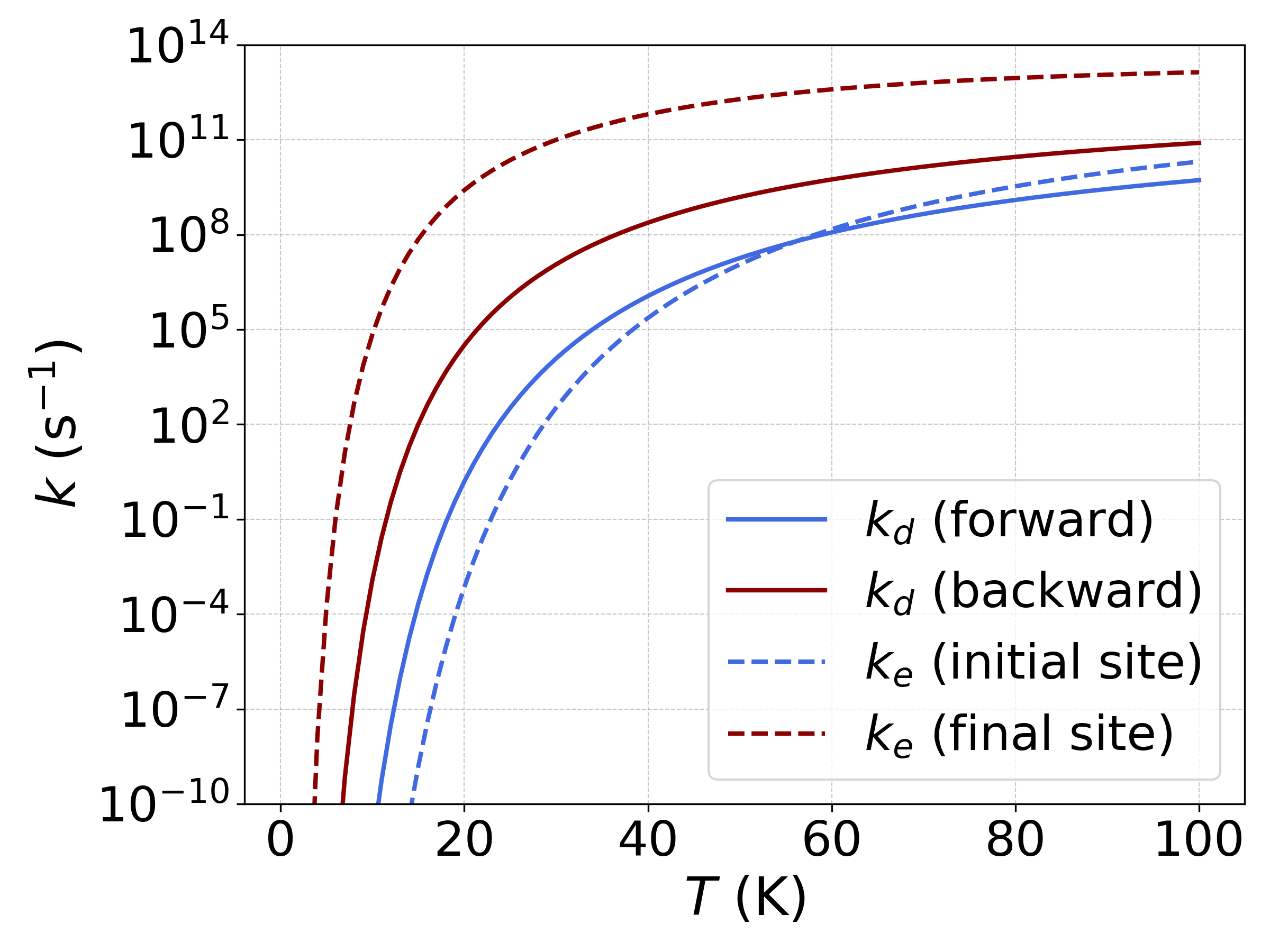}
\caption{}
\label{subfig:diff_vs_des_3}
\end{subfigure}
\hfill
\begin{subfigure}[b]{0.46\linewidth}
\includegraphics[width=\linewidth]{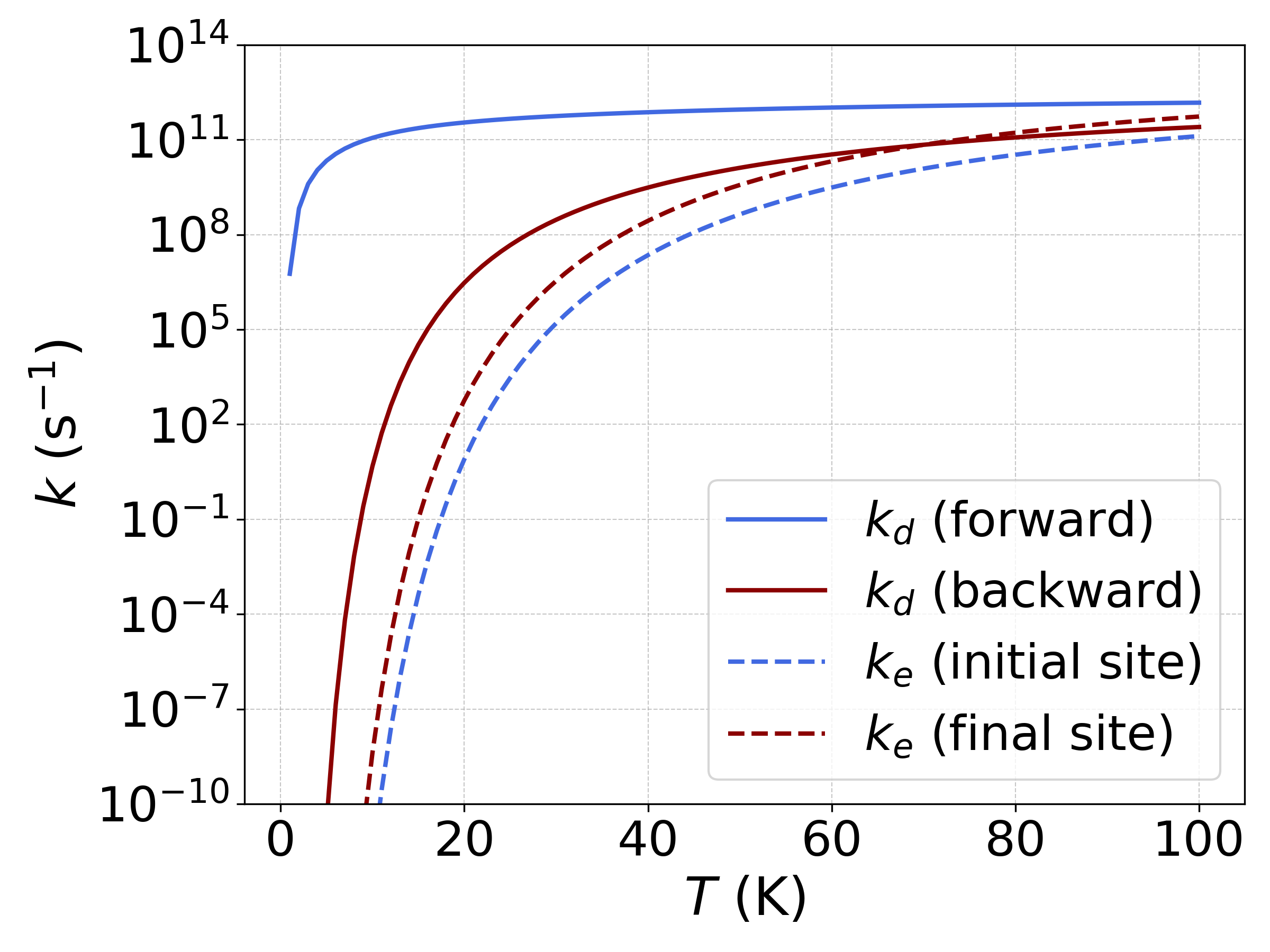}
\caption{}
\label{subfig:diff_vs_des_4}
\end{subfigure}

\caption{\label{fig:rate_comparison}Comparison between diffusion ($k_{\text{d}}$) and desorption ($k_{\text{e}}$) rate coefficients for 4 representative site pairs as a function of temperature.  Full lines refer to the diffusion coefficients (blue for the forward diffusion from the initial to the final site, and red for the backward diffusion from the final to the initial site). The dotted lines indicate the desorption coefficients of the 2 sites involved in the diffusive event (dotted blue lines refer to desorption from the initial site, while the dotted red lines refer to the desorption from the final site).}
\end{figure}

In Fig.~\ref{fig:rate_comparison}, we report 4 representative examples (for the same representative site pairs shown in Fig.~\ref{fig:meps}) of rate coefficient profiles with respect to the temperature, for the forward and reverse diffusion processes, and for the desorption process.

At low temperatures, diffusion appears to be systematically faster than desorption and, by increasing the temperature, the two processes start to happen on similar time scales. This behavior reflects the general trend observed for the distribution in Fig~\ref{fig:kd_vs_ke}.

\section{Gaussian fit accuracy}
\label{app_C}
The accuracy of the bidimensional Gaussian fits used to approximate the KDE distributions was quantitatively assessed by computing the relative error between the fitted surface and the corresponding KDE distribution.
At each temperature between 10 and 100~K, the relative error between them was computed at each grid point $(k_\mathrm{d},k_\mathrm{e})$ with the relation
\begin{equation}
    \varepsilon (k_\mathrm{d},k_\mathrm{e}) = \frac{\big|f_{\mathrm{fit}}(k_\mathrm{d},k_\mathrm{e}) - f_{\mathrm{KDE}}(k_\mathrm{d},k_\mathrm{e})\big|}{f_{\mathrm{KDE}}(k_\mathrm{d},k_\mathrm{e})} \,,
\end{equation}
where $f_{\mathrm{KDE}} (k_\mathrm{d},k_\mathrm{e})$ and $f_{\mathrm{fit}} (k_\mathrm{d},k_\mathrm{e})$ represent the values of the two functions at the same coordinates.
To avoid numerical instabilities in low-probability regions, only the points satisfying the relation $f_{\mathrm{fit}} \geq 0.1\,f_{\mathrm{fit}} ^{\mathrm{max}}$ are included in the analysis, restricting the comparison to the most significant region of the distribution. Across the explored temperature range (10$-$100~K), the median relative error was found to lie between 0.17 and 0.26, while the 25th and 75th percentiles of the error distribution were in the 0.08$-$0.11 and 0.30$-$0.63 ranges, respectively.

\section{Astrophysical framework}
\label{app_D}
We employ the Minimum Mass Solar Nebula (MMSN) model, with the following density and temperature radial profiles, assuming hydrostatic equilibrium and vertical isothermality,

\begin{equation}
    \rho(r) = \mu m_{\mathrm p}n(r) = \frac{\Sigma_0}{H(r)\sqrt{2\pi}}\bigg( \frac{r}{1\hspace{0.2em}\mathrm{au}} \bigg) ^{-3/2}
\end{equation}
\begin{equation}
\label{T_radial_profile}
    T_{\mathrm d}(r) = T(r) = T_0 \bigg( \frac{r}{1\hspace{0.2em}\mathrm{au}} \bigg) ^{-1/2} \,,
\end{equation}
where $r$ represents the distance from the central star, $m_\mathrm{p}$ is the proton mass, $\mu=2.34$ the approximated mean molecular weight, and a disk vertical scale height
\begin{equation}
    H(r) = \frac{c_{\mathrm s}}{\Omega_\mathrm K} \,,
\end{equation}
where the speed of sound and the Keplerian angular frequency, respectively, are given by
\begin{equation}
    c_\mathrm s = \sqrt{\frac{k_\mathrm B T(r)}{\mu m_\mathrm p}} \hspace{2em} \text{and} \hspace{2em} \Omega_\mathrm K = \sqrt{\frac{GM_*}{r^3}} \,,
\end{equation}
with $G$ being the gravitational constant and assuming the following reference values: $M_* = 1\rm M_{\odot}$ (solar mass), $\Sigma_0 = 1700$ g\,cm$^{-2}$, $T_0 = 200$ K.

\end{appendices}

\end{document}